\documentclass[twocolumn,apj]{openjournal}
\usepackage{graphicx,amsmath,amssymb,amstext}
\usepackage{relsize}
\usepackage{amsbsy,amsfonts,amsthm,color,bm}
\usepackage{placeins}
\usepackage[colorlinks,linkcolor=blue,citecolor=blue,urlcolor=blue ]{hyperref}
\usepackage{cleveref}
\usepackage[utf8]{inputenc}
\usepackage[caption=false]{subfig}
\usepackage{float}

\begin{document} 

\title{Tomographic halo model of the unWISE-Blue galaxies using cross-correlations with BOSS CMASS galaxies}

\author{Alex Krolewski}
\affiliation{Waterloo Centre for Astrophysics, University of Waterloo, Waterloo, ON N2L 3G1, Canada}
\affiliation{Department of Physics and Astronomy, University of Waterloo, Waterloo, ON N2L 3G1, Canada}
\affiliation{Perimeter Institute for Theoretical Physics, Waterloo, ON N2L 2Y5, Canada}
\email{akrolews@uwaterloo.ca}

\author{Jensen Lawrence}
\affiliation{Waterloo Centre for Astrophysics, University of Waterloo, Waterloo, ON N2L 3G1, Canada}
\affiliation{Department of Physics and Astronomy, University of Waterloo, Waterloo, ON N2L 3G1, Canada}
\affiliation{Department of Earth, Atmospheric and Planetary Sciences, Massachusetts Institute of Technology, Cambridge, MA 02139, USA}

\author{Will J.~Percival}
\affiliation{Waterloo Centre for Astrophysics, University of Waterloo, Waterloo, ON N2L 3G1, Canada}
\affiliation{Department of Physics and Astronomy, University of Waterloo, Waterloo, ON N2L 3G1, Canada}
\affiliation{Perimeter Institute for Theoretical Physics, Waterloo, ON N2L 2Y5, Canada}




\begin{abstract}
The halo model offers a framework for investigating galaxy clustering, and for understanding the growth of galaxies and the distribution of galaxies of different types. Here, we use the halo model to study the small-scale
clustering and halo occupation distribution (HOD) of the unWISE-Blue galaxy sample, an infrared-selected sample of $\sim$100 million galaxies across the entire extragalactic sky at $z\sim 0.5$
-- similar redshifts to the Baryon Oscillation Spectroscopic Survey (BOSS) CMASS sample. Although the photometric unWISE galaxies cannot be easily split in redshift, we use their cross-correlation with the BOSS CMASS sample to tomographically probe the HOD of the unWISE galaxies at $0.45 < z < 0.75$. To do so, we develop a new method for applying the halo model to cross-correlations between a photometric sample and a spectroscopic sample in narrow redshift bins, incorporating halo exclusion, post-Limber corrections, and redshift-space distortions. 
We reveal strong evolution in the CMASS HOD, and modest evolution in the unWISE-Blue HOD. For unWISE-Blue, we find that the average bias and mean halo mass drop from $b = 1.6$ and $\log_{10}(M_{\mathrm{h}}/M_{\odot}) \sim 13.4$ at $z \sim 0.5$ to $b = 1.4$ and $\log_{10}(M_{\mathrm{h}}/M_{\odot}) \sim 13.1$ at $z \sim 0.7$, and that the satellite fraction drops modestly from $\sim$20\% to $\sim$10\% in the same range.
These results are useful for creating mock samples of the unWISE-Blue galaxies. Furthermore, the techniques developed to obtain these results are applicable to other tomographic cross-correlations between photometric samples and narrowly-binned spectroscopic samples, such as clustering redshifts.
\end{abstract}

\section{Introduction}

Over a period of 10 years, the Wide-field Infrared Survey Explorer satellite \citep[WISE;][]{Wright2010} mapped the sky in the infrared, initially in four wavebands for the first year with active cooling, and then in two bands, at 3.4 and 4.6 $\mu$m, for the remaining nine years \citep[the NEOWISE post-cryogenic survey;][]{Mainzer2011, Mainzer2014}.
This survey has produced several very large catalogs of infrared sources, such as the unWISE catalog of 2 billion sources from the first five years of observations \citep{Schlafly2019}.
A simple infrared color cut and the use of Gaia data to reject bright point sources  yields a sample of 500 million galaxies at $0 < z < 2$ across the entire sky.
The high number density and full sky coverage make unWISE an ideal sample for cross-correlations with other full-sky tracers, e.g.\ CMB secondary anisotropies
such as lensing \citep{Krolewski2019, Krolewski2021, Kusiak22,Farren23b,Farren23}, the kinetic Sunyaev-Zel'dovich effect \citep{Kusiak21, Bloch24}, the integrated Sachs-Wolfe effect \citep{KrolewskiFerraro22}, and other full-sky samples like high-energy neutrinos \citep{Ouellette24} and the cosmic infrared backgpround \citep{Yan24}.
To make further use of this sample, we must clearly understand the types of galaxies it contains.

To this end we employ the halo model, a framework for understanding the distribution of galaxies within the large-scale cosmic web \citep{Seljak2000, Ma2000, Peacock2000, Cooray2002, Smith2002}. In the halo model, galaxies exist only within halos of dark matter, where each halo may have at most one central galaxy and an arbitrary number of satellite galaxies. The distribution of central and satellite galaxies within a halo, and the distribution of halos that host each type of galaxy, are controlled by the halo occupation distribution \citep[HOD;][]{Berlind2003,Zehavi2005,Zheng2007}. Different HOD types are required for different galaxy types. Here, we adopt a strict halo model where halos are assumed to be spherical, and we use analytic expressions or numerical fits for the halo density profile, halo mass function, matter power spectrum, halo bias, concentration-mass relation and an exclusion model to account for overlaps between halos. The details are given in Section~\ref{sec:hod}.

By fitting the parameters of the halo model to the clustering of the unWISE sample, we can determine the basic properties of these galaxies within the halo model framework. The unWISE galaxies are split into three samples at $z\sim0.5$, $1.1$, and $1.5$ using a simple color cut \citep{Schlafly2019,Krolewski2019}. 
As a first step towards modelling their HOD, we consider the lowest-redshift ``Blue'' sample, with a broad redshift distribution centered at $z\sim0.5$ and extending to $z \sim 1$. The sample includes galaxies across a wide range in redshift, but further subdivision in $z$ is not possible because the two-band WISE photometry cannot be used to construct accurate photometric redshifts.
Cross-matching to optical imaging surveys
would improve this situation,
but would considerably reduce the sky coverage and decrease the number density, both of which are
undesirable for many cross-correlation applications.

In order to understand the tomographic clustering, we consider the cross-correlation with the spectroscopic Sloan Digital Sky Survey \citep[SDSS;][]{York00} Baryon Oscillation Spectroscopic Survey \citep[BOSS;][]{Dawson13} CMASS galaxies (see Section~\ref{sec:cmass} for details of this sample). By dividing CMASS into many narrow redshift bins and cross-correlating with unWISE, we can measure
unWISE clustering tomographically, without requiring photometric redshifts
for the unWISE galaxies. This approach
is conceptually similar to clustering redshifts \citep{Newman08, McQuinn13, Menard13}.

This allows us to go beyond the limits of a photometric sample, enhancing our ability to understand the evolution of the sample with redshift. 
We describe the unWISE and CMASS datasets used in Section~\ref{sec:data}.
The correlation and cross-correlation measurements are presented in Section~\ref{sec:corr}. 
Next, the extensions to the halo model required to probe the cross-correlation are described in Section~\ref{sec:hod-model}. 
Finally, the resulting HOD constraints are presented in Section~\ref{sec:results}. As a by-product of our analysis we fit the CMASS galaxies, matching previous results \citep{White2011, Alam17, Reid14, Saito16}. For the unWISE galaxies, we find evidence for a strongly evolving HOD, and we are able to compare the host halos of unWISE and CMASS galaxies. These results are discussed further in our conclusions, summarized in Section~\ref{sec:conclusions}. 


\section{Data}
\label{sec:data}

\subsection{unWISE}  \label{sec:WISE}

The WISE satellite mapped the entire sky at 3.4 (W1), 4.6 (W2), 12 (W3), and 22 (W4) $\mu$m for one year in the main mission \citep{Wright2010}, and subsequently for nine additional years in W1 and W2 for the non-cryogenic post-hibernation NEOWISE phase \citep{Mainzer2011,Mainzer2014}. 
This deep all-sky infrared imaging is ideal for identifying large samples of $z \gtrsim 1$ galaxies: the unWISE catalog \citep{Lang2014,Meisner2017} contains 500 million galaxies across the entire sky.
With a simple W1-W2 color cut, these galaxies can be divided into three samples at different redshifts, the Blue, Green, and Red samples at $z \sim 0.5$, 1.1, and 1.5, respectively. These samples are extensively described in \citet{Krolewski2019} and \citet{Schlafly2019},
and their properties are summarized in Table 1 of \citet{Krolewski2019}.
Stars are removed by matching to Gaia DR2 \citep{Gaia_DR2} and removing any unWISE
source within $2.75''$ (one WISE pixel) of a Gaia point source.
Residual stellar contamination is $<2\%$,
as estimated from cross-matching to the COSMOS field with high-resolution multi-band optical imaging \citep{Krolewski2019}. We directly measure $s$, the slope of the number counts as a function of magnitude $m$, $s \equiv d\log_{10}(N/\mathrm{d}m)$, by perturbing the photometry of the unWISE galaxies and applying the same color cuts \citep[see Appendix D of][]{Krolewski2019}.

\begin{figure}
\centering
\includegraphics[width=0.48\textwidth]{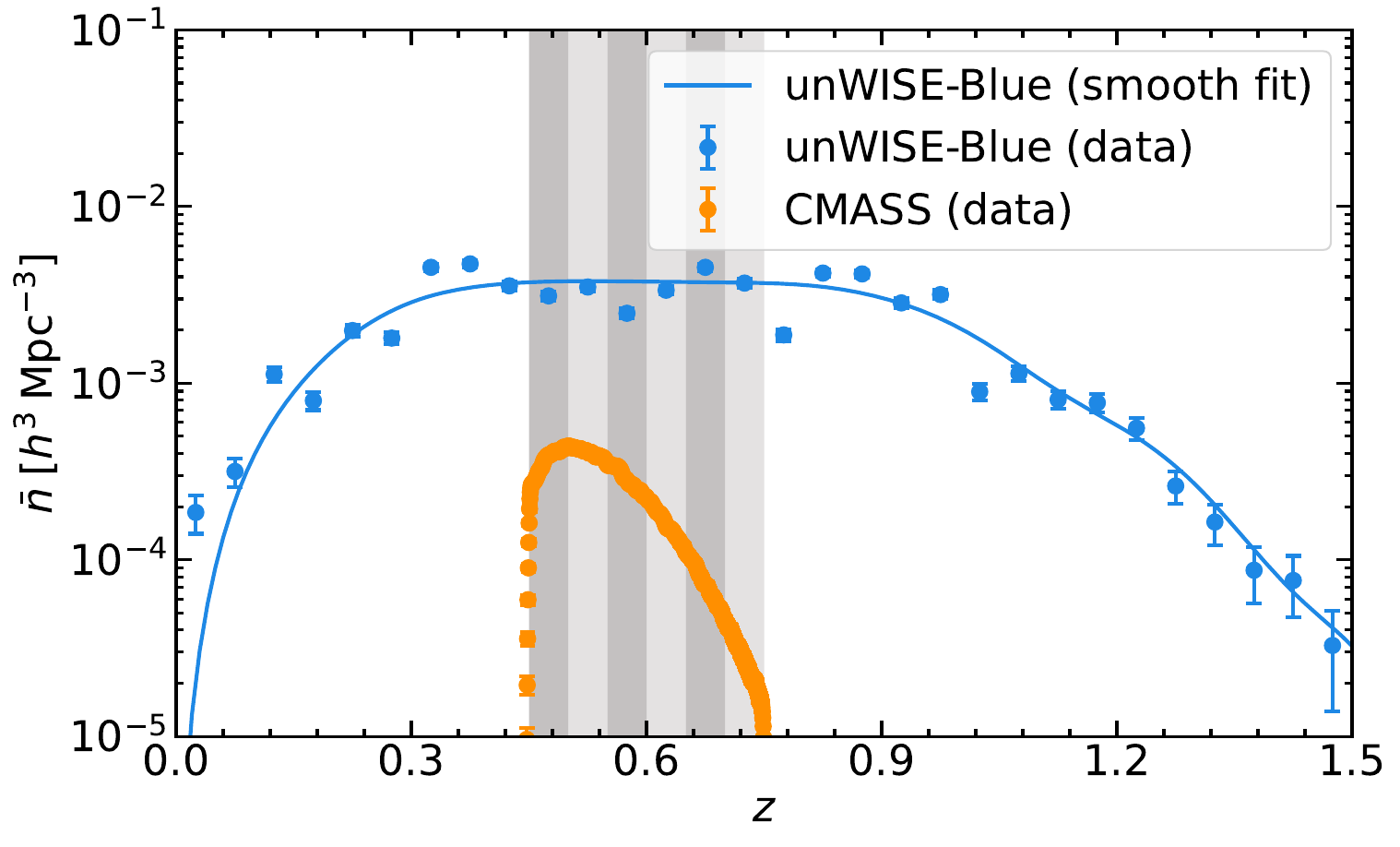}
    \caption{Number densities of unWISE-Blue (blue) and CMASS (orange) galaxy samples. The gray shaded regions demarcate the different redshift bins. Uncertainties on the unWISE-Blue $\bar{n}(z)$ are from Poisson fluctuations in the observed photo-$z$ number counts in the 2 deg$^2$ COSMOS field.}
        \label{fig:dndz}
\end{figure}

We measure the unWISE redshift distribution using the COSMOS2015
photometric redshifts in the 2 deg$^2$ COSMOS field \citep[Fig.~\ref{fig:dndz};][]{Laigle2015}.
These redshifts are determined from SED-fitting to galaxy photometry across many deep optical and near-infrared bands, using a best-fit galaxy or AGN template.
We use the photometric redshift from the nearest COSMOS source within $2.75''$, first removing any COSMOS source fainter than 20.7 (19.2) Vega magnitude at 3.6 (4.2) $\mu$m. This is necessary
to eliminate spurious matches caused by blends in WISE, which has a $6''$ PSF.

To represent the photometric redshift, the COSMOS2015 catalog reports the median of the redshift likelihood given a galaxy template. It also reports a best-fit likelihood from an AGN fit. Because these infrared-bright unWISE sources could have AGN contribution, we use the redshift for the AGN fit if it provides a better fit to the multi-band photometry, and otherwise use the median of the likelihood for the galaxy template.
AGN contamination is low, especially for the Blue sample that we fit in this work, both because AGN mid-infrared colors are different from the Blue color cuts,
and because the Gaia point-source cut removes the brightest quasars.
Ultimately,
only 19\% of the COSMOS matches are better-fit by an AGN SED, and using the AGN redshifts makes minimal difference on the redshift distribution.

unWISE-Blue has a broad redshift distribution peaking around $z \sim 0.6$ but extending from $z \sim 0$ to $z \sim 1$ (Fig.~\ref{fig:dndz}). In this work, we focus on constraining its halo occupation distribution near the peak of the redshift distribution at $0.45 < z < 0.75$, where we have a well-characterized, large-area spectroscopic sample, SDSS-BOSS CMASS. We focus on this cross-correlation as a first demonstration of the method and leave the cross-correlation with higher and lower redshifts spectroscopic samples (to fully measure the evolving HOD of unWISE-Blue) to future work.

We additionally apply the linear imaging systematics weights described
in \citet{KrolewskiFerraro22} and \citet{Farren23} to correct
for spurious correlations between unWISE galaxy density and WISE depth and stellar density. After applying these weights, the unWISE
galaxy density is uncorrelated with nine imaging systematics templates
(WISE depth, extinction, dust density, and DIRBE Sky and Zodi Atlas \citep[DSZA; ][]{Kelsall98}\footnote{DSZA data are publicly available at \url{https://lambda.gsfc.nasa.gov/product/cobe/dirbe_dsza_data_get.cfm}} estimates
of diffuse infrared sky background).

\subsection{CMASS}  \label{sec:cmass}

The Baryon Oscillation Spectroscopic Survey \citep[BOSS;][]{Dawson13,Eisenstein11} targeted
1.5 million galaxies, stars and quasars selected from 10,000 deg$^2$ of SDSS DR8 imaging \citep{Gunn98,York00,Gunn06,Fukugita96,Lupton01,Smith02,Pier03,Padmanabhan08,Doi10,Aihara11}.
The upgraded BOSS spectrograph \citep{Smee2013} enabled highly accurate
and complete redshifts for these targets.
The majority of the targets are luminous red galaxies at $0.2 < z < 0.75$, composing the CMASS and LOWZ samples. Successful redshifts
are assembled into a uniformly-selected large-scale structure catalog in \cite{Reid16}.

In this paper, we use the CMASS galaxies as a spectroscopic tracer
to tomographically probe the halo occupation distribution of the unWISE
galaxies across a broad range of redshifts.
We therefore measure the cross-correlation between unWISE galaxies and six bins from CMASS with $\Delta z = 0.05$ from $z=0.45$ to 0.75.
We restrict the clustering measurement to the union of the CMASS LSS mask
and the unWISE mask.\footnote{CMASS catalogs and masks are publicly available at \url{https://data.sdss.org/sas/dr12/boss/lss/}}
Around 75\% of the CMASS sample is located in the North Galactic Cap (NGC).
As the sample is slightly different in the South Galactic Cap (SGC), we
restrict our measurement to the NGC for simplicity; the additional constraining power from SGC is minimal, and may require different
HOD parameters due to the slightly different selection.

Like unWISE, the CMASS galaxy density must be corrected for its dependence on imaging systematics.
It is also affected by fiber collisions: no galaxy pairs with separation $<55$'' can be observed due to the physical size of the BOSS
spectrograph fibers.
Galaxies lost to fiber collisions tend to occupy
overdense environments; therefore, accurately measuring
the clustering requires correcting for fiber collisions.
Like the imaging systematics, fiber collisions
are approximately mitigated in the catalog by creating another
set of weights, upweighting the nearest neighbor
to a collided galaxy. We apply the default
weights in the catalog,
\begin{equation}
    w_{\textrm{tot}} = w_{\textrm{systot}}
    (w_{\textrm{cp}} + w_{\textrm{noz}} - 1).
\end{equation}
where $w_{\textrm{systot}}$ corrects for imaging systematics, $w_{\textrm{cp}}$ is the close-pair weight accounting for fiber collisions, and $w_{\textrm{noz}}$ corrects for redshift failures.
We also restrict our autocorrelation
measurements to scales larger than the 55'' collision scale (corresponding to 0.3 $h^{-1}$ Mpc at $z = 0.45$ and 0.5 $h^{-1}$ Mpc at $z = 0.75$).

Finally, due to the broad redshift kernel of the unWISE galaxies, we must account
for magnification of CMASS galaxies by foreground unWISE galaxies, which depends
on the number counts slope $s$ for CMASS. We determine $s$ by directly measuring the response of the galaxy number counts to a change in magnitude: we add a uniform offset to the
 CMASS galaxies' flux across all bands,\footnote{This is only correct for fluxes measured with an infinitely large aperture. Lensing magnification preserves surface brightness and increases the size of the galaxies. If the aperture is small compared to the galaxy's size, then the flux increase from magnification is reduced since some of the magnified galaxy is shifted outside the fixed aperture \citep{WenzlChen23,Zhou23}. We correctly account for this effect using the composite deVaucoleurs-exponential model in the SDSS cmodel magnitudes.} and also account for the change in redshift success rate as the galaxies get brighter from lensing magnification using the scaling of success rate with flux from Fig.~7 of \cite{Reid16}.
We then apply the CMASS targeting color cuts to the shifted photometry.
Measurements of $s$ for CMASS are given in Appendix D of \cite{Farren23},
and are very similar to the measurements in \cite{WenzlChen23}.
We report the magnification bias measurements in Table~\ref{tab:magnification_cmass}, and use $s = 0.453$ for the unWISE-Blue sample, as measured in \cite{Krolewski2019}.

\begin{table}
\centering
\begin{tabular}{c|c|c}
\toprule
$z_{\textrm{min}}$ & $z_{\textrm{max}}$ & $s$ \\
\hline
0.45 & 0.50 & 0.937 \\
0.50 & 0.55 & 1.040 \\
0.55 & 0.60 & 1.148 \\
0.60 & 0.65 & 1.352 \\
0.65 & 0.70 & 1.553 \\
0.70 & 0.75 & 1.957 \\
\hline
\end{tabular}
\caption{Magnification bias coefficient $s$ for the different CMASS redshift bins, reproducing Table 13 in \citet{Farren23}. We use $s = 0.453$ for the unWISE-Blue sample.}
\label{tab:magnification_cmass}
\end{table}

\section{CMASS-unWISE cross-correlation measurement}
\label{sec:corr}

We use the Davis-Peebles estimator\footnote{The high number density of the unWISE-Blue sample makes the computational cost of the RR term in the Landy-Szalay estimator prohibitive.} 
\citep{DavisPeebles},
\begin{equation}
    w_{sp}(\theta) = \frac{D_s D_p}{D_s R_p} \frac{N_R}{N_D} - 1,
\end{equation}
where $s$ and $p$ indicate the spectroscopic and photometric samples, respectively,
to compute the cross-correlation in 10 log-spaced bins between 0.1 and 10 $h^{-1}$ Mpc at the mean redshift
of each bin. We use a tree-based correlation function code to store the
shared unWISE data and random catalogs across all redshift bins.\footnote{\url{https://github.com/akrolewski/BallTreeXcorrZ}}
The unWISE randoms are distributed within the unWISE mask,
with the expected number of randoms in each pixel scaled using the sub-pixel
mask accounting for area lost within each NSIDE=2048 HEALPix \citep{Gorski05} pixel due to diffraction spikes, latents, or ghosts around modestly bright WISE sources.

\begin{figure}
\includegraphics[width=0.49\textwidth]{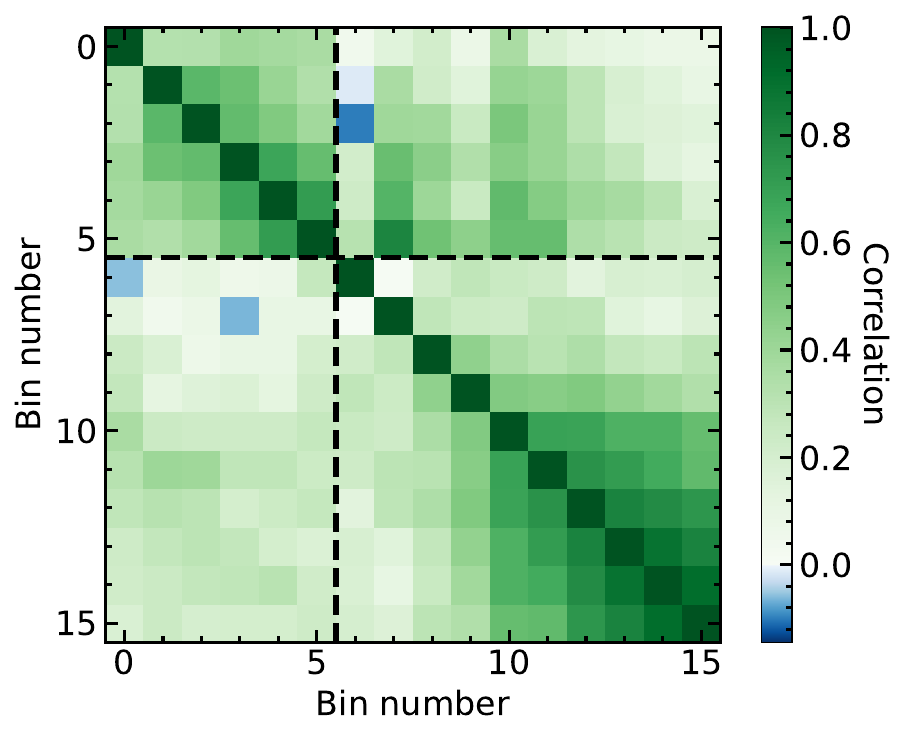}
    \caption{The jackknife correlation matrix for the $0.5 < z < 0.55$ redshift bin, with axes labelled by bin number. The black dotted lines separate the matrix into its component blocks: the upper left block is the CMASS autocorrelation matrix while the lower right block is the CMASS-unWISE cross-correlation matrix. The autocorrelation matrix has fewer elements due to the small-scale cut removing scales affected by fiber collisions.}
    \label{fig:cov}
\end{figure}

We use jackknife resampling to compute the covariance matrix (see \cite{MohammadPercival} for discussion on the accuracy of the jackknife covariance).
We create jackknife regions by splitting the sky into NSIDE=8 HEALPix pixels, and then joining neighboring pixels until the regions are roughly the same size.
This produces 136 jackknife regions.
The error on $w_{12}(\theta)$ becomes
\begin{equation}
    \sigma^2_w(\theta) = \sum_{L=1}^N \frac{R_{[L]}}{R} \left(w_{[L]}(\theta) - w(\theta)\right)^2,
\end{equation}
where $R$ refers to the numbers of randoms, the subscript $[L]$ indicates that we exclude the $L$th region (correcting the conventional jackknife factor for the fact that the regions are slightly different sizes; see Equation 5 in \citet{Myers05}), and $w(\theta)$ is simply the measured
cross-correlation using the full sky. We use the same jackknife regions for both the auto and cross-correlations,
allowing us to compute the full covariance matrix between all spatial bins and between the auto and cross-correlations (Fig.~\ref{fig:cov}).

Due to the finite number of jackknife realizations, there is a sample variance associated
with the covariance matrix. This noise biases the expectation value of the inverse covariance matrix
\citep{Hartlap_2006,DodelsonSchneider,Percival14,SellentinHeavens16} and changes the form of the likelihood \citep{Percival22}. We use the Gaussian approximation of \cite{Percival22}, following their Equations 54--56 to modify the covariance matrix.

The correlation function is normalized using the galaxy density across the entire survey area.
However, the survey area is finite, meaning galaxy density across the survey area is not necessarily equal to the true galaxy density across the entire Universe.
This can be corrected by subtracting the integral constraint, calculated by \citep{RocheEales,Coil12}
\begin{equation}
    \mathrm{IC} = \frac{1}{\Omega^2} \iint w(\theta) \; \mathrm{d}\Omega_1 \: \mathrm{d}\Omega_2,
\end{equation}
where $\Omega$ is the area of the survey and the integrals run over the survey window function.
To study the size of the integral constraint, we compute it for a single example redshift bin, $0.45 < z < 0.50$, using CAMB \citep{Lewis:1999bs,Howlett:2012mh} to compute $w(\theta)$ (out to very large scales) from the nonlinear power spectrum times fiducial linear biases of 1.87 and 1.68 appropriate for unWISE-Blue and CMASS. We find the integral constraint is 1.6\% of the cross-correlation in the largest bin ($8 < r < 10$ $h^{-1}$ Mpc) and negligible compared to the much larger CMASS autocorrelation. This is equal to 0.3$\sigma$ for the final bin cross-correlation, i.e.\ considerably smaller than the statistical error. Due to the small size of the integral constraint,
we neglect it hereafter.

\section{The HOD model}  \label{sec:hod-model}
In this section, we describe the HOD model used to jointly characterize the CMASS autocorrelation and CMASS-unWISE cross-correlation data. We begin by taking a step back and examining halo model described in \citet{Halomod20} (see also the recent review in \cite{Asgari24}), whose \texttt{HALOMOD} code we use and modify.

\subsection{Halo model}  \label{sec:hod}

The halo model states that at large scales, all matter is bound to ``halos'' of dark matter. It is assumed that each halo is spherically symmetric and that halo shape depends solely on halo mass. This implies that the halo density is strictly a function of halo mass and distance from the halo center, i.e. $\rho_{\mathrm{h}} = \rho_{\mathrm{h}}(r, M)$. Therefore, the density field in a region of the Universe that contains $n$ halos is
\begin{equation}
    \rho(\bm{x}) = \sum_{i = 1}^n \rho_{\mathrm{h}}(|\bm{x} - \bm{x}_i|, M_i), \label{eq:matter-density}
\end{equation}
where $\bm{x}_i$ and $M_i$ are the position of the center and mass, respectively, of the $i$th halo. However, this function is difficult to compute in practice. The power of the halo model lies in its ability to convert Equation~(\ref{eq:matter-density}) into a semi-analytic form that can be readily calculated computationally. To do so, the halo model requires the following pieces of information:
\begin{itemize}
    \item The moments of the halo density profiles ($\langle \rho_{\mathrm{h}}(r, M) \rangle$, $\langle \rho_{\mathrm{h}}^2(r, M) \rangle$, etc.).
    We use the Navarro-Frenk-White profile \citep{NFW}.
    \item The halo mass function (HMF) $n(M)$, which provides the abundance of halos of a given mass. We use the \cite{Tinker08} mass function. Masses are reported in their default units, spherical overdensity masses with $\Delta = 200$ relative to the matter density.
    \item The matter overdensity
    \begin{equation}
        \delta(\bm{x}) \equiv \frac{\rho(\bm{x}) - \bar{\rho}}{\bar{\rho}}, \label{eq:overdensity}
    \end{equation}
    where $\bar{\rho}$ is the mean of $\rho(\bm{x})$. This is set by the cosmology and primordial power spectrum;
    we use a Planck 2015 \citep{Planck15} cosmology with $\sigma_8 = 0.831$, $n_s = 0.9645$, and the nonlinear
    power spectrum from \citet{Takahashi2012}.
    \item The halo bias $b_{\mathrm{h}}(\bm{x}, M)$, which is the expected overdensity of halos of mass $M$ in a region with matter overdensity $\delta(\bm{x})$. We use the mass-bias relation of \cite{Tinker10},
    and additionally allow for a scale-dependent bias following \cite{TinkerWeinberg05}.
    \item The concentration-mass relation $c(M)$, which describes the link between halo shape and halo mass. We use the formulation of \cite{Duffy08}.
    \item A ``halo exclusion'' model to account for double counting of correlations within halos in both the 1-halo and 2-halo terms of the halo model. This is necessary to accurately model the transition region between the 1-halo and 2-halo terms, both of which are relevant to the scales of this work. Halo exclusion has previously been difficult to model. Using the NgMatched model of \cite{TinkerWeinberg05}, we model exclusion between triaxial ellipsoids with a range of axis ratios. This method is cumbersome because it requires double integrals over both halo mass distributions at every wavenumber; it is simplified by calculating the number density with 2D integrals
    and then matching the number density in separable 1D integrals elsewhere.
\end{itemize}

\subsection{HOD for cross-correlations}

Thus far, the framework of the halo model considers only the bulk properties of halos, meaning each halo is essentially featureless beyond its density profile. To add galaxies to the halo model, we must specify $N(M)$, the HOD function that describes the number of galaxies that populate a halo of mass $M$.

In this paper, we use the HOD model developed in \cite{Zheng2007}, which states that
\begin{equation}
    \langle N(M) \rangle = \langle N_{\mathrm{c}}(M) \rangle + \langle N_{\mathrm{s}}(M) \rangle, \label{eq:Ntot}
\end{equation}
where
\begin{equation}
    \langle N_{\mathrm{c}}(M) \rangle = \frac{1}{2}\left[ 1 + \mathrm{erf}\left( \frac{\log M - \log M_{\mathrm{min}}}{\sigma_{\log M}} \right) \right] \label{eq:Ncen}
\end{equation}
and
\begin{equation}
    \langle N_{\mathrm{s}}(M) \rangle = \langle N_{\mathrm{c}}(M) \rangle \left( \frac{M - M_0}{M_1} \right)^{\alpha}, \label{eq:Nsat}
\end{equation}
where $\log$ is the base-10 logarithm ($\log_{10}$). $\langle N_{\mathrm{c}}(M) \rangle$ and $\langle N_{\mathrm{s}}(M) \rangle$ are the mean number of central and satellite galaxies, respectively, that one expects to find in a halo of mass $M$. $\langle N_{\mathrm{c}}(M) \rangle$ follows a binomial distribution and ranges between 0 and 1 (since there can be at most one central galaxy), while $\langle N_{\mathrm{s}}(M) \rangle$ follows a Poisson distribution and can be arbitrarily large.

The HOD parameters $M_{\mathrm{min}}$, $M_0$, $M_1$, $\alpha$, and $\sigma_{\log M}$ control the relationship between halo mass and galaxy number. Specifically \citep{Zheng2007},
\begin{itemize}
    \item $M_{\mathrm{min}}$ is the characteristic minimum mass a halo must have to host a central galaxy;
    \item $M_0$ sets a cutoff below which one finds zero satellites;
    \item $M_1$ sets the mass at which one finds one satellite per halo;
    \item $\alpha$ is the slope of the satellite galaxy number;
    \item $\sigma_{\log M}$ is the width of the central galaxy cutoff profile.
\end{itemize}


Moreover, we assume the ``central condition'' for the CMASS HOD, which
requires that halos cannot host a satellite galaxy without also hosting a central. This is physically
reasonable because central galaxies are typically more luminous than satellite galaxies, so if the satellite is bright enough to be included in the survey, then the central must be too.
We do not assume the central condition for the unWISE-Blue HOD, since unWISE has a bright cut ($\mathrm{W1} > 15.5$) that removes a fairly large number of galaxies; hence, the satellite could be in the sample, but the central
would be too bright. Practically, this makes very little difference, since the data favors HODs
where $N_{\mathrm{s}}$ drops off at much higher masses than $N_{\mathrm{c}}$. The central condition is not compatible
with a binomial $N_{\mathrm{c}}$ and a Poissonian $N_{\mathrm{s}}$ \citep{Beutler13}; the Poisson distribution for $N_{\mathrm{s}}$ is therefore modified
to satisfy the central condition as shown in Table 3 of \cite{Asgari24}.

When considering cross-correlations, we must also consider the correlations between centrals and satellites of the two galaxy populations: if one sample inhabits a halo,
is the other sample more or less likely to also inhabit that halo? These correlations are parameterized
by $R_{\mathrm{cs}}$, $R_{\mathrm{ss}}$, and $R_{\mathrm{sc}}$, which are correlation coefficients controlling correlations
between centrals and satellites of each type \citep{Miyaji11,Krause12}. For simplicity, we fix all three parameters to zero (the uncorrelated case); exploring the impact of these parameters is left for future work.


The \cite{Zheng2007} model assumes that at high mass, every halo hosts a central galaxy ($N_{\mathrm{c}}(M) = 1$).
However, CMASS has color cuts and  flux limits creating incompleteness: the color cuts can reduce completeness
at high mass while the flux limits generally reduce completeness at low mass and high redshift.
The impact of flux limits is studied in \citet{Zhai17} and \citet{Tinker17}, who find that the \cite{Zheng2007} HOD can reproduce the real galaxy occupation very well at low mass, largely by changing $M_{\textrm{min}}$ and $\sigma_{\log M}$. The impact of color cuts is studied in \citet{Leauthaud16}, who directly quantify
CMASS completeness by comparing the CMASS stellar mass function to the stellar mass function of a massive
galaxy sample selected without color cuts. At $z < 0.6$, color cuts remove galaxies from CMASS even at the massive end, causing $N_{\mathrm{c}}(M)$ to asymptote to 30--50\% below unity. This can be modelled by uniformly reducing $N(M)$ by a scaling factor $f_{\textrm{inc}}$, the asymptotic completeness at the high stellar-mass end.
\cite{Leauthaud16} report $f_{\textrm{inc}}$ in several redshift bins across the CMASS range; we reproduce
their calculation using our exact redshift bins, using their publicly available data \citep{Bundy15}.\footnote{\url{https://www.ucolick.org/~kbundy/massivegalaxies/s82-mgc.html}}
Specifically, we linearly interpolate their double-Schechter function fits for the total stellar mass function
(Table 1 of \cite{Leauthaud16}); measure the CMASS stellar mass function in a given redshift bin using their Stripe 82 catalog and masses, requiring a match to CMASS; and determine $f_{\textrm{inc}}$ by fitting the completeness $c$ as a function of $M_\star$ by (their Equation 17):
\begin{equation}
    c = \frac{f_{\textrm{inc}}}{2} \left[1 + \mathrm{erf} \left( \frac{\log(M_*/M_1)}{\sigma} \right)\right].
\end{equation}
with free parameters $f_{\textrm{inc}}$, $M_1$, and $\sigma$.
We then apply this $f_{\textrm{inc}}$ scaling factor to $N_{\mathrm{c}}(M)$ in our implementation of the halo model.

Likewise, it is also possible that the unWISE sample does not include
every central galaxy, i.e.\ $N_{\mathrm{c}} < 1$ at high mass.
unWISE also includes a bright cut, which could remove high-mass halos.
Unlike CMASS, detailed studies such as \cite{Leauthaud16} have not been performed, so the high-mass incompleteness is unknown.
To encompass a range of possibilities, we will consider two limiting cases:
in one (the default case), we require that the model matches the observed unWISE $\bar{n}(z)$ within the uncertainty (hence $N_{\textrm{c}} = 1$ at high mass), and in the other, we impose
the observed $\bar{n}(z)$ as a lower bound on the model $\bar{n}(z)$ (hence $N_{\textrm{c}} < 1$ at high mass), equivalent to marginalizing over $f_{\textrm{inc}}$ with an uninformative prior. More complicated possibilities
are certainly possible, for instance a Gaussian distribution for $N_{\textrm{c}}$ \citep[e.g.][]{Rocher23}, but they are beyond the scope of this paper, in which we only consider the \cite{Zheng2007} model as a first step
in modelling the unWISE-Blue HOD. Within this model, the two cases
considered span the uncertainty in unWISE incompleteness.

\subsection{Angular cross-correlations}
\label{sec:hod-cross}

With the specifics of the HOD established, we turn our attention to angular correlations. Considering two galaxy fields, hereafter labelled 1 and 2., the two-point angular correlation function for these fields is given by \citep[e.g.][]{Halomod20}
\begin{equation}
    w_{12}(\theta) \approx \int_0^{\infty}  \int_0^{\infty} p(r_1)p(r_2)\xi_{12}(R, \bar{r}) \: \mathrm{d}r_1 \: \mathrm{d}r_2, \label{eq:w12-original}
\end{equation}
where
\begin{equation}
    p_i(r) \equiv \frac{\mathrm{d} N_i}{\mathrm{d} r}
\end{equation}
is the probability density of finding a galaxy from the field $i$ at a line-of-sight comoving distance of $r$, $\xi_{12}$ is the two-point spatial correlation function for the fields 1 and 2,
\begin{equation}
    R \equiv \sqrt{r_1^2 + r_2^2 - 2r_1r_2\cos\theta}
\end{equation}
is the projected separation between the points $r_1$ and $r_2$, and
\begin{equation}
    \bar{r} \equiv \frac{r_1 + r_2}{2}
\end{equation}
is the the mean line-of-sight comoving distance. 

\subsubsection{Beyond the Limber approximation \label{sec:beyond-limber}}  

To facilitate numerical calculations of $w_{12}$, Equation \ref{eq:w12-original} is often approximated to a more computationally tractable form. For many surveys, each $p_i(r)$ is a broad distribution in redshift, and $\theta$ is a small angle, meaning it is appropriate to employ the approximation developed in \cite{Limber1953}. However, Limber's equation diverges for narrow galaxy bins and wide angles \citep{Simon07}. While we work on small scales in this paper, our spectroscopic bins are very narrow, with $\Delta z = 0.05$. Hence, the Limber approximation is insufficient, and we apply the post-Limber formulation developed by \cite{Simon07}:
\begin{multline}
    w_{12}(\theta) \approx \frac{1}{1 + \cos\theta} \; \mathlarger{\times} \\ \int_0^{\infty} 
    \int_{\sqrt{2(1 - \cos\theta)}\bar{r}}^{2\bar{r}} Q(\bar{r}, \Delta) 
    \xi_{12}(R, \bar{r}, \mu) \frac{R}{\Delta} \: \mathrm{d}R \: \mathrm{d}\bar{r}, \label{eq:w12-simon}
\end{multline}
where
\begin{equation}
    \Delta \equiv \frac{1}{\sqrt{2}}\sqrt{\frac{R^2 - 2\bar{r}^2(1 - \cos\theta)}{1 + \cos\theta}}
\end{equation}
and
\begin{equation}
    Q(\bar{r}, \Delta) \equiv p_1(\bar{r} - \Delta)p_2(\bar{r} + \Delta) + p_1(\bar{r} + \Delta)p_2(\bar{r} - \Delta),
\end{equation}
while $\mu$ is the cosine of the angle between $R$ and $\bar{r}$.

\subsubsection{Redshift-space distortions \label{sec:rsd}}

We also include the effect of redshift-space distortions (RSD) in the model by allowing $\xi_{12}$ to be a function of separation $R$, angle $\mu$, and comoving distance $\bar{r}$ (or equivalently redshift).
The impact of RSD on angular clustering is small \citep{Krywonos24}, so it is appropriate to implement an approximate linear RSD model \citep{Kaiser87} where $\xi_{12}$ in Equation \ref{eq:w12-simon} takes the form 
\begin{equation}
    \xi_{12}(R, \bar{r}, \mu) = \sum_{\ell \in \{0, 2, 4\}} \alpha_\ell(\beta_1, \beta_2) \xi_{12}^{\ell}(R, \bar{r}) L_{\ell}(\mu),
\end{equation}
where the RSD parameter $\beta \equiv f/b$ and $f$ is the growth factor, commonly approximated as $\Omega_m(z)^{0.55}$.
Allowing for the cross-correlation of two tracers with differing biases, the coefficients $\alpha$ are:
\begin{align}
    \alpha_0 &= 1 + \frac{1}{3}(\beta_1 + \beta_2) + \frac{1}{5}\beta_1 \beta_2, \\
    \alpha_2 &= \frac{2}{3}(\beta_1 + \beta_2) + \frac{4}{7}\beta_1 \beta_2, \\
    \alpha_4 &= \frac{8}{35}\beta_1 \beta_2,
\end{align}
and the multipole moments of the correlation function are
\begin{equation}
    \xi_{\ell}(R, \bar{r}) = \frac{i^\ell}{2\pi^2} \int_0^\infty P_{12}(k, \bar{r} )j_{\ell}(kR) k^2 \: \mathrm{d}k.
    \label{eqn:xiell}
\end{equation}

\begin{figure*}
\includegraphics[width=1.0\textwidth]{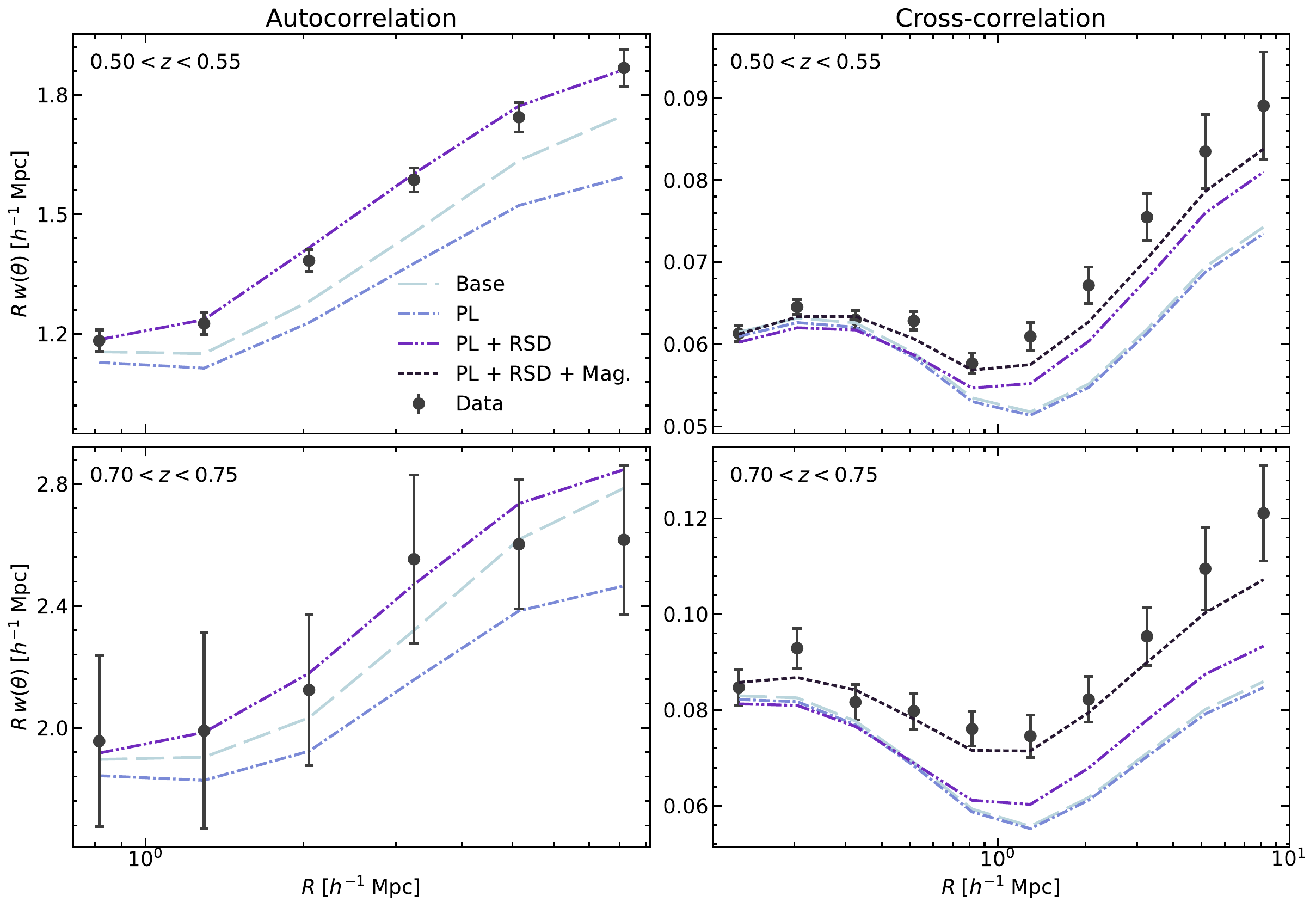}
    \caption{The angular correlation function data and best-fit models for the CMASS autocorrelation (left column) and the CMASS-unWISE cross-correlation (right column) in the redshift bins $0.5 < z < 0.55$ (top row) and $0.7 < z < 0.75$ (bottom row), with jackknife errors as described in Section~\ref{sec:corr}. The Base best-fit considers only the standard Limber approximation to $w(\theta)$. In contrast, the PL best-fit uses the post-Limber equation described in Section \ref{sec:beyond-limber}. Next, the PL + RSD best-fit adds the effects of redshift-space distortion explained in Section \ref{sec:rsd}. Finally, the PL + RSD + Mag. best-fit includes the magnification bias from Section \ref{sec:magnification}. 
    Since magnification is negligible for the autocorrelation, we take PL + RSD as the overall best-fit model for the CMASS autocorrelation, and PL + RSD + Mag. as the overall best-fit model for the CMASS-unWISE cross-correlation.
    }
    \label{fig:model}
\end{figure*}

\subsubsection{3D power spectrum: one-halo and two-halo terms}

Next, we define the model for the 3D power spectrum $P_{12}(k, \bar{r})$ in Equation~\ref{eqn:xiell}.
Within the framework of the halo model, it is convenient to break the power spectrum into a term accounting for intrahalo galaxy correlations (the 1-halo term) and a term accounting for interhalo galaxy correlations (the 2-halo term). In particular, this decomposition is best done in real space, defining $P_{12}(k)$ as the Fourier transform of $\xi_{12}(r)$ (and dropping the redshift label $\bar{r}$ for clarity):
\begin{align}
    P_{12}(k) &= 4 \pi \int_0^{\infty} \xi_{12}(r)r^2j_0(kr) \: \mathrm{d}r, \\
    \xi_{12}(r) &= [1 + \xi_{12}^{1\mathrm{h}}(r)] + \xi_{12}^{2\mathrm{h}}(r),
\end{align}
where $j_0(x) = \sin(x)/x$ is the zeroth-order spherical Bessel function.
The 1-halo term is given by
\begin{equation}
    \xi_{12}^{\mathrm{1h}}(r) = \frac{1}{2\pi^2}\int_0^{\infty}  P_{12}^{\mathrm{1h}}(k)k^2j_0(kr) \: \mathrm{d}k,
\end{equation}
where
\begin{multline}
    P_{12}^{\mathrm{1h}}(k) = \int n(M)u(k|M) \Big{[}N_{\mathrm{s},1}N_{\mathrm{s},2}u(k|M) \\ + \langle N_{\mathrm{c},1}N_{\mathrm{s},2} \rangle + \langle N_{\mathrm{c},2}N_{\mathrm{s},1} \rangle \Big{] \: \mathrm{d}M}
    \label{eqn:p121h}
\end{multline}
is the 1-halo power spectrum (generalizing Eq.~63 in \cite{Halomod20}), with $u(k|M)$ being the mass-normalised Fourier transform of the halo profile.
In this model, the self-pair shot-noise term is ignored, and it can be added back to the power spectrum to match the observed $\bar{n}$. In the correlation function shot noise creates a delta-function at $r=0$ and thus does not impact observables with a finite scale cut.
However, Eq.~\ref{eqn:p121h} creates a spurious shot-noise power on large scales, which allows the 1-halo term to dominate the 2-halo term on very large scales. We use \texttt{HALOMOD}'s \texttt{force\_1halo\_turnover} option to remove this term.

The 2-halo term is given by
\begin{equation}
    \xi_{12}^{\mathrm{2h}}(r) = \frac{1}{2\pi^2}\int_0^{\infty}  P_{12}^{\mathrm{2h}}(k)k^2j_0(kr) \: \mathrm{d}k,
\end{equation}
where the 2-halo power spectrum is
\begin{multline}
    P_{\mathrm{12}}^{2\mathrm{h}}(k) = \iint I_{\mathrm{g}}(k,M_1)I_{\mathrm{g}}(k, M_2) \times \\ b(k,M_2)b(M_1,r)b(M_2,r)P_{\mathrm{m}}(k) \: \mathrm{d}M_1 \: \mathrm{d}M_2
\end{multline}
and
\begin{equation}
I_g(k, M) = \frac{1}{\bar{n}_g} n(M) u(k | M) \langle N(M) \rangle.
\end{equation}
from Eqs.~15 and 62 in \cite{Halomod20}.
Note that the $M_1$ and $M_2$ integrals are separable following the simplifications of the ``Ng-Matched'' exclusion model (as the triaxial exclusion model otherwise couples the mass limits) as described above in Section~\ref{sec:hod}.

\subsubsection{Magnification bias \label{sec:magnification}}

Because unWISE has a very broad galaxy kernel, the cross-correlation arises both from clustered
objects at the same redshift, and lensing magnification correlating galaxies at different redshifts:
\begin{equation}
    w_{12}^{\textrm{mag}}(\theta) = w_{\mu1g2}(\theta) + w_{g1\mu^2}(\theta) + w_{\mu1\mu2}(\theta)
\end{equation}
where the magnification contribution arises from galaxy magnification of each sample correlated with galaxies from the other sample ($w_{g \mu}(\theta)$), and the cross-correlation between galaxy magnification of the two samples  ($w_{\mu1 \mu2}(\theta)$).

We separately compute the magnification contribution to the angular cross-correlation and add it to the clustering term, using the magnification bias values in Table~\ref{tab:magnification_cmass}.
Since the magnification term is subdominant, for simplicity we do not model it with the halo model,
but rather with a simple linear bias times the nonlinear \texttt{HALOFit} \citep{Mead20} power spectrum, requiring that the linear bias
matches the linear bias of the halo model considered at each point in the MCMC chain. We compute $w_{12}^{\textrm{mag}}(\theta)$ by working with the angular power spectrum $C_\ell$ (where $\ell$ are spherical harmonic multipoles) and transforming it to the angular correlation function:
\begin{equation}
w_{XY}(\theta)= \sum_\ell \frac{2 \ell + 1}{4\pi}C^{XY}_\ell P_\ell(\cos{\theta)}
\end{equation}
where $P_\ell$ are the Legendre polynomials.
Working within the Limber approximation (valid due to the smallness of the non-Limber magnification terms),
\begin{equation}
    C_\ell^{XY} = \int_0^{\chi_\star} d\chi \frac{W^X(\chi)W^Y(\chi)}{\chi^2}P_{\textrm{mm}}(k \chi = \ell + 1/2)
\end{equation}
where $\chi$ is the comoving distance and $P_{\textrm{mm}}(k)$ is the nonlinear matter power spectrum. The galaxy kernel is given by
\begin{equation}
    W^{g,i}(\chi) = b(z(\chi)) \frac{dN_i}{d\chi}
\end{equation}
where $b(z)$ matches the linear bias of the halo model. Finally, the magnification kernel is a function of $s$ given by
\begin{equation}
    W^{\mu,i}(\chi) = (5s_{\mu}-2)\frac{3}{2}\Omega_mH_0^2(1+z)g_i(\chi)
\end{equation}
\begin{equation}
    g_i(\chi)=\int_{\chi}^{\chi_{\ast}} d\chi' \frac{\chi(\chi'-\chi)}{\chi'}H(z')\frac{dN_i}{dz'}
\end{equation}

\subsubsection{Binning the angular cross-correlation}

We measure the angular cross-correlation as a function of physical scale by converting
the angle to a transverse distance using the comoving distance to the mean spectroscopic
redshift of each bin, converting $w_{12}(\theta)$ to $w_{12}(R)$. We then compute the binned cross-correlation to compare to the data, averaging $w_{12}(R)$ within each bin in $R$ according to
\begin{equation}
    w_{12,\textrm{bin},i} = \frac{1}{R_{\textrm{max},i}^2 - R_{\textrm{min},i}^2} \int_{R_{\textrm{min},i}}^{R_{\textrm{max},i}} 2 R \, w_{12}(R) \: \mathrm{d}R.
\end{equation}

The results of this model model are shown for two example bins at the low and high redshift end in Fig.~\ref{fig:model},
broken into its constituent terms to demonstrate the effects of the post-Limber, RSD, and magnificantion terms.

\subsection{Likelihood}

We then compute the standard Gaussian likelihood
\begin{equation}
    \mathcal{L}_{\textrm{clus}} = -\frac{1}{2} (\bm{w}_{\textrm{data}} - \bm{w}_{\textrm{mod}}) \mathbf{C}^{-1}(\bm{w}_{\textrm{data}} - \bm{w}_{\textrm{mod}})^T
\end{equation}
where the data vectors run over both CMASS autocorrelation and CMASS-unWISE cross-correlation.
We also add two terms of the form
\begin{equation}
    \mathcal{L}_{\bar{n}} = -\frac{1}{2} \left(\frac{\bar{n}_{\textrm{mod}} - \bar{n}_{\textrm{data}}}{\sigma_{\bar{n}}}\right)^2
\end{equation}
to the likelihood for the number density of the CMASS and unWISE galaxies. When marginalizing over the unWISE $f_{\textrm{inc}}$, we do not change the clustering
(as this parameter only affects the total number of galaxies),  but replace the Gaussian likelihood in $\bar{n}$ with a step function imposing the constraint that $\bar{n}_{\textrm{mod}} > $ $\bar{n}_{\textrm{data}}$.

We sample from the likelihood using the \texttt{Cobaya} MCMC sampler \citep{cobaya}, sampling until the errorbars
estimated from many sequential chains change by less than 10\% when adding a new chain.
We impose flat priors on the 10 HOD parameters (five for the CMASS HOD and five for the unWISE-Blue HOD), chosen to be large enough to never be informative.

\begin{figure*}
\centering
\includegraphics[width=1.0\textwidth]{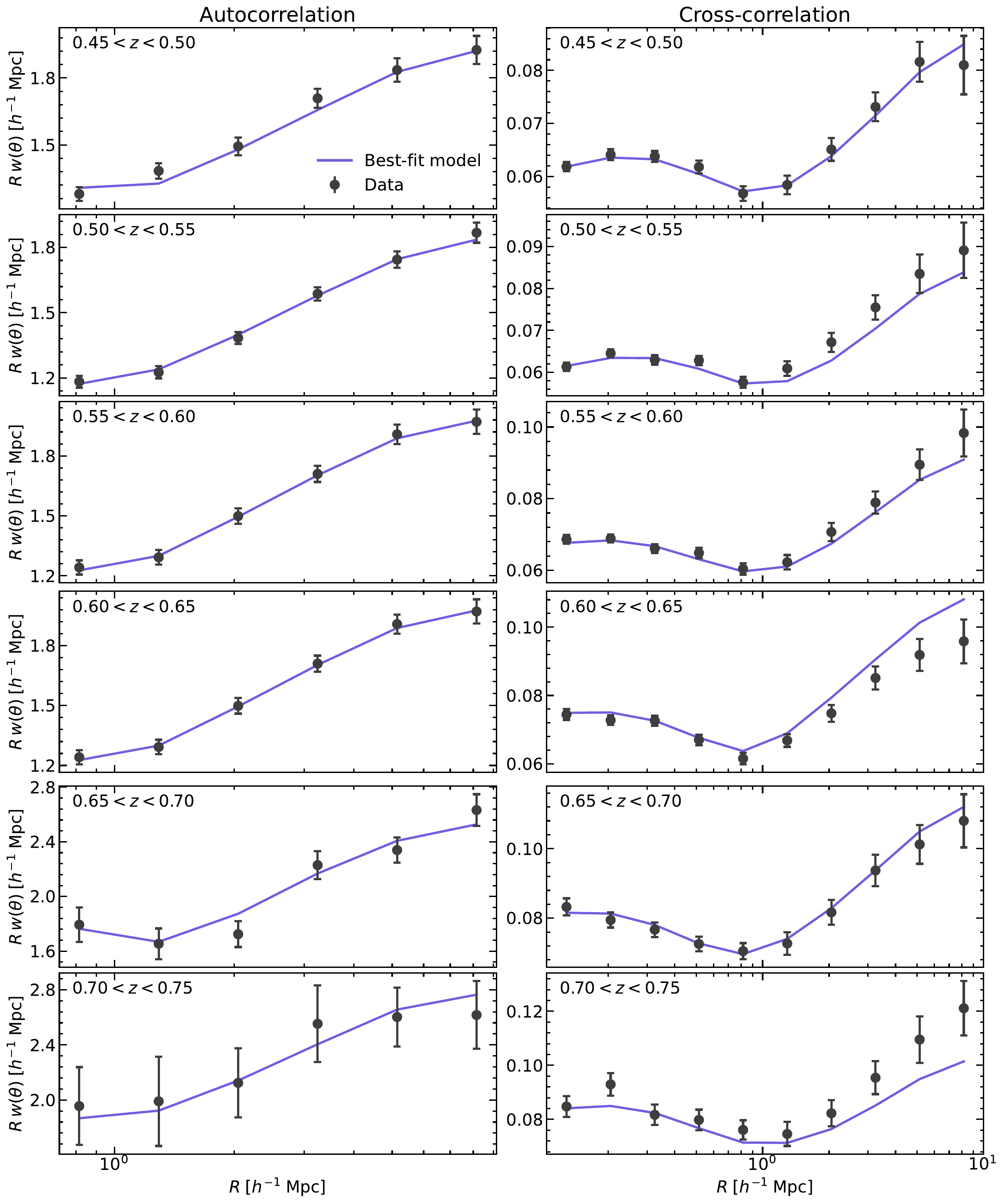}
    \caption{The angular correlation function data and overall best-fit models for the CMASS autocorrelation (left column) and the CMASS-unWISE cross-correlation (right column). From top to bottom, the rows correspond to the redshift bins $0.45 < z < 0.5$, $0.5 < z < 0.55$, $0.55 < z < 0.6$, $0.6 < z < 0.65$, $0.65 < z < 0.7$, and $0.7 < z < 0.75$. The model fits the data well in both autocorrelation and cross-correlation across all redshift bins.}
    \label{fig:best_fit_models}
\end{figure*}

\section{HOD constraints}  \label{sec:results}

We show the best-fit and median marginalized results,
and their errorbars (16th to 84th percentile range)
for all HOD parameters and derived parameters in
Tables~\ref{tab:results_cmass} and~\ref{tab:results_wise},
for the baseline case where unWISE-Blue is assumed to be fully complete ($f_{\textrm{inc}} = 1$).
We plot the best-fit models in Fig.~\ref{fig:best_fit_models}. The model fits the data well in all redshift bins (with 9 degrees of freedom).

\begin{figure*}
\centering
\includegraphics[width=1.0\textwidth]{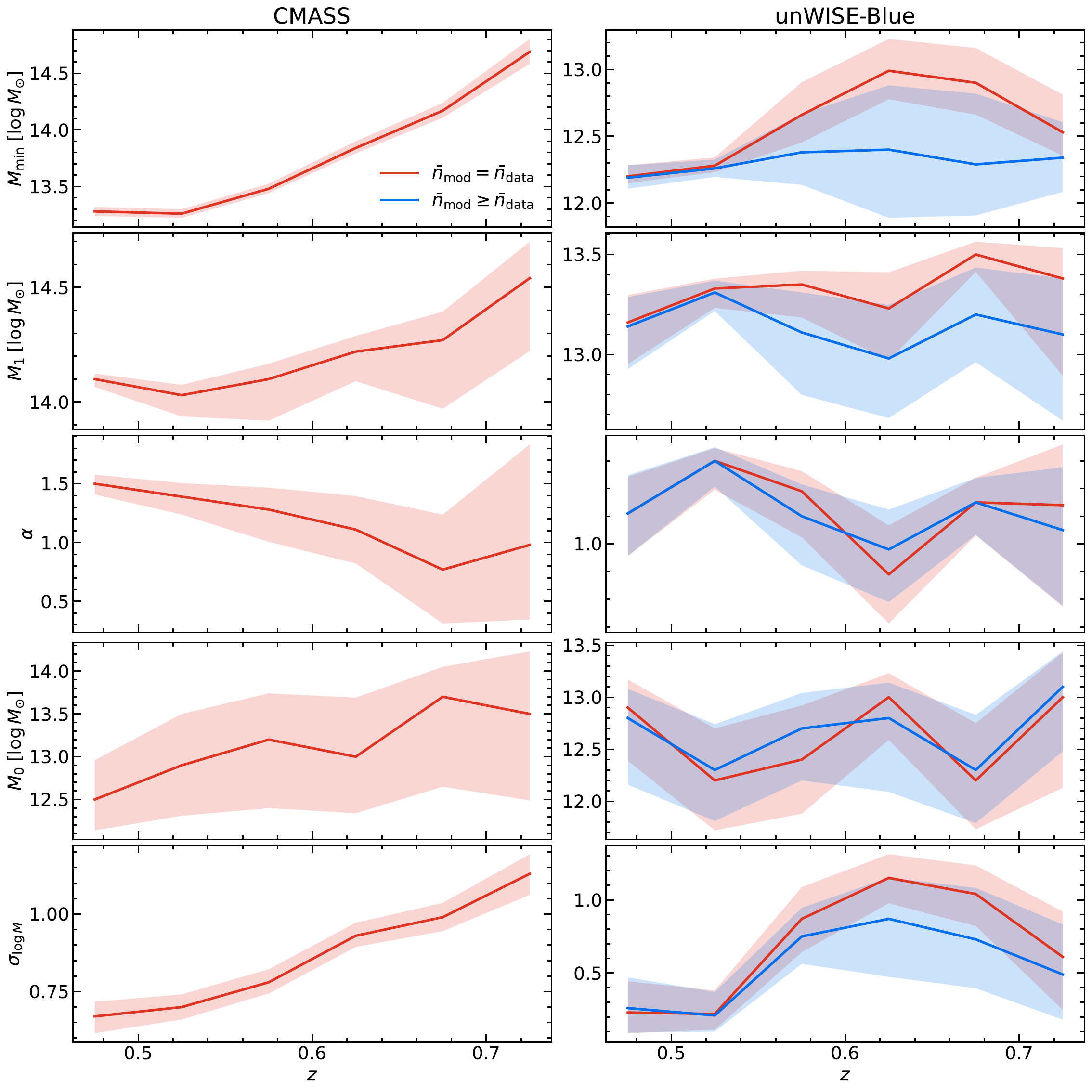}
\caption{Evolution of the 5 HOD parameters for the CMASS (left column) and unWISE-Blue (right column) galaxies as a function of redshift. The unWISE-Blue panels show both the default case where we use the observed $\bar{n}$ in the likelihood, and the ``incompleteness'' case where the model is allowed to have higher number density than the data (corresponding to a situation where the unWISE-Blue galaxies randomly down-sample best-fit HOD). The shaded regions represent the $\pm 1\sigma$ uncertainties derived from the MCMC samples.}
\label{fig:evolution}
\end{figure*}

\begin{figure*}
\centering
\includegraphics[width=1.0\textwidth]{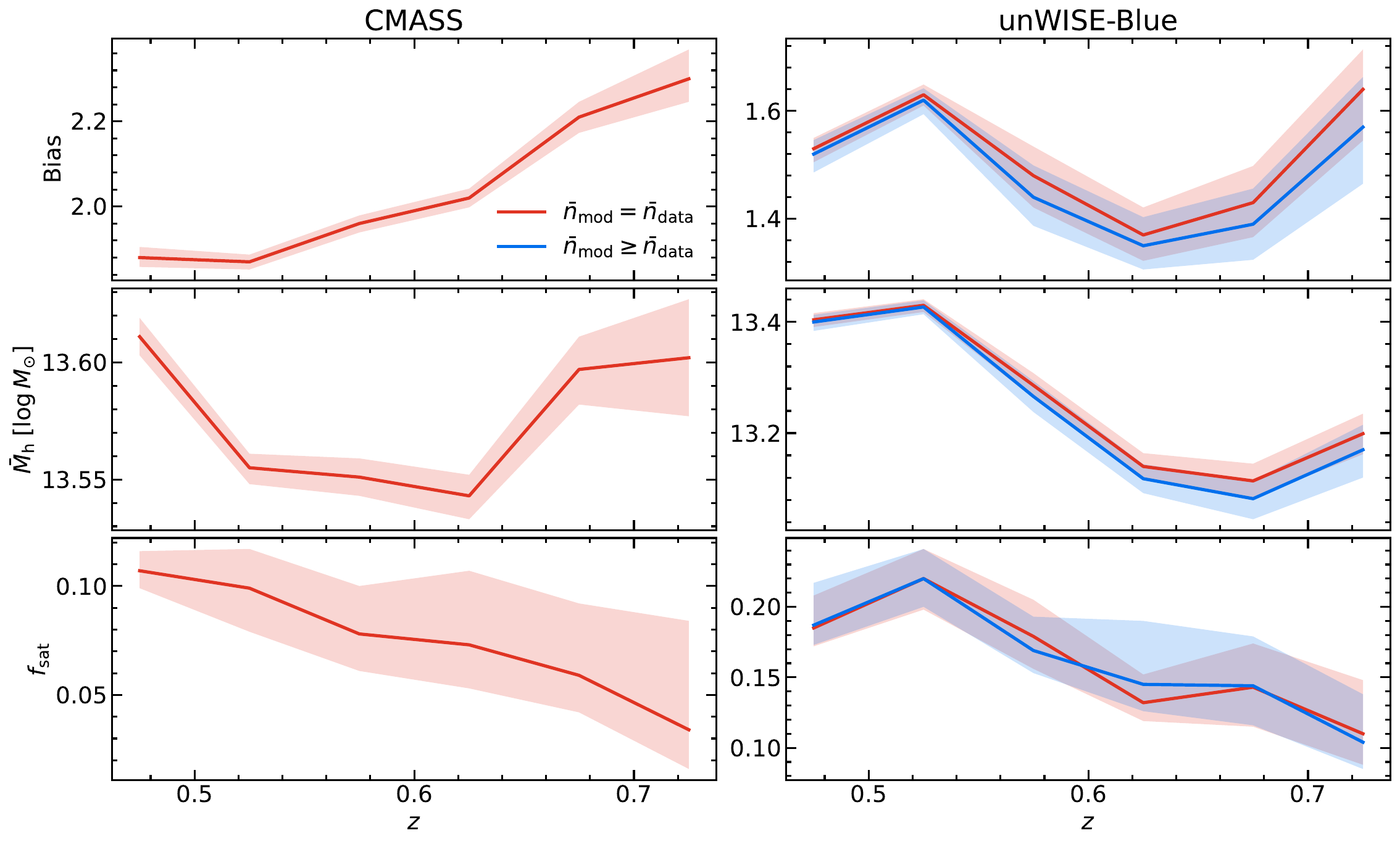}
\caption{Evolution of the derived bias, mean halo mass ($\bar{M}_h)$, and satellite fraction ($f_{\mathrm{sat}}$), parameters for the CMASS (left column) and unWISE-Blue (right column) galaxies as a function of redshift. The unWISE-Blue panels show both the default case where we use the observed $\bar{n}$ in the likelihood, and the ``incompleteness'' case where the model is allowed to have higher number density than the data (corresponding to a situation where the unWISE-Blue galaxies randomly down-sample best-fit HOD). The shaded regions represent the $\pm 1\sigma$ uncertainties derived from the MCMC samples.}
\label{fig:evolution2}
\end{figure*}

We plot the evolution of the parameters in Figs.~\ref{fig:evolution} and~\ref{fig:evolution2}.
We show the constraints on the HOD parameters in Fig.~\ref{fig:contours} and derived parameters in Fig.~\ref{fig:contours_derived} for a single representative redshift bin at $0.55 < z < 0.60$.
The strong degeneracy directions are very similar
between all redshift bins, though the size of the contours
increases as the correlation functions
become noisier at higher redshift. We therefore only
show contours from a single bin for clarity.
In this redshift bin, the constraints when requiring the model to match the observed unWISE $\bar{n}$
are considerably stronger than when the observed $\bar{n}$ is only treated as a lower bound.
On the other hand, in the first two and last redshift bins, the constraints are very similar to each other regardless of what we assume for the unWISE incompleteness.

\subsection{CMASS}

\subsubsection{Redshift evolution of the CMASS HOD}

\begin{figure}
\includegraphics[width=0.47\textwidth]{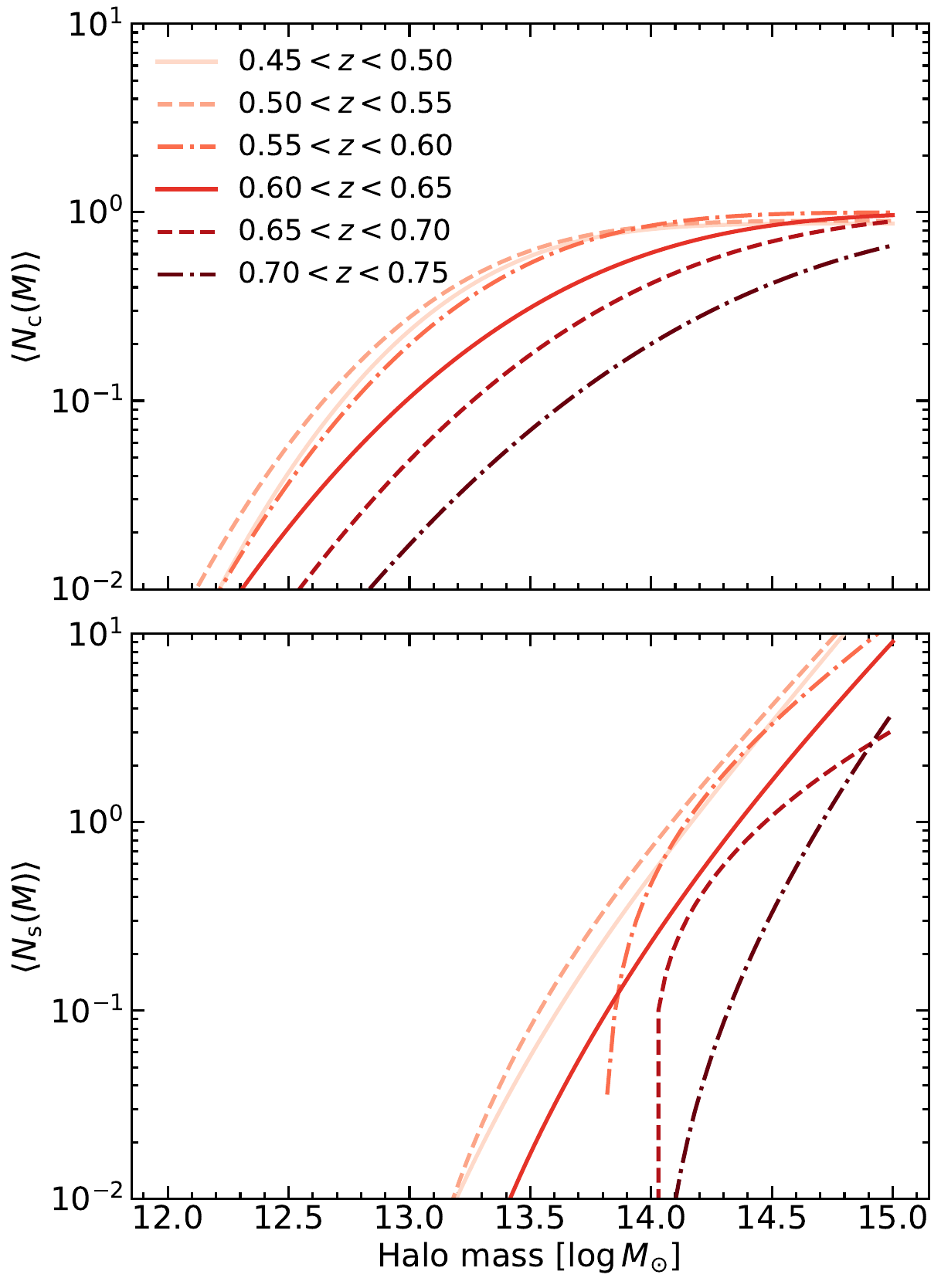}
    \caption{The mean number of central galaxies (top panel) and the mean number of satellite galaxies (bottom panel) for the CMASS HOD as functions of halo mass over all redshift bins. Note the evolution of both $\langle N_{\mathrm{c}}(M) \rangle$ and $\langle N_{\mathrm{s}}(M) \rangle$ as a function of redshift. The sharp cutoff in $\langle N_{\mathrm{s}}(M) \rangle$
    for $0.55 < z < 0.60$ and $0.60 < z < 0.65$ is due to the larger values of $\log_{10}(M_0/M_\odot)$ (the mass at which there are zero satellite galaxies) for these bins, 13.8 and 14.0 respectively.
    \label{fig:cmass_hod_evolution}}
\end{figure}

For CMASS, $M_{\textrm{min}}$ and $\sigma_{\log{M}}$ both significantly increase with redshift.
This leads to an evolving HOD, i.e.\ the best-fit
CMASS HOD in a single bin cannot fit the data in any other bin (holding the unWISE-Blue HOD constant).
The evolving $N(M)$ is shown in Fig.~\ref{fig:cmass_hod_evolution}.
Despite the varying HOD parameters,
the mean halo mass is constant from $0.5 < z < 0.65$ at $\log_{10}(M_{\mathrm{h}}/M_{\odot}) \sim 13.55$, with a slight (but statistically significant) increase at $z > 0.65$, and a similarly small decrease in the first redshift bin.
This is due to a cancellation between the changes to $\langle N(M) \rangle$ (which favor higher halo mass at higher redshift)
and the halo mass function, which exponentially decreases at the high mass
end toward higher redshift.
A halo of fixed mass is more biased at higher redshift, causing the 
mean halo bias to increase with redshift.
The increase in $M_{\textrm{min}}$ and $\sigma_{\log{M}}$ also decreases the number density by decreasing $\langle N(M) \rangle$.
Within the errorbars, $b(z) D(z)$ is quite close to constant at $z < 0.65$ (ranging from 1.42 to 1.47) but then rises to 1.56 and 1.62 in the last two redshift bins. Finally, the satellite fraction declines with increasing redshift,
but the changes are not statistically significant; the satellite fraction is $10^{+1.8}_{-1.7}\%$ in the $0.5 < z < 0.55$ bin.

\subsubsection{Comparison to past work}


These results for the CMASS HOD
are similar to past findings.
\cite{White2011} fit an HOD model to the
correlation function measured from 44,000
CMASS galaxies at $0.4 < z < 0.7$ observed in the first semester
of BOSS. Measuring
within a single broad redshift bin
at $z_{\textrm{eff}} = 0.57$,
they find $M_{\textrm{min}} = 13.08 \pm 0.12$, $M_1 = 14.06 \pm 0.10$, $\sigma_{\log{M}} = 0.60 \pm 0.15$,
$\alpha = 0.90 \pm 0.19$, and $M_0 = 13.13 \pm 0.16$.
These results are similar to our results
in the $0.5 < z < 0.55$ and $0.55 < z < 0.6$ bin,
though with a lower $M_{\textrm{min}}$ and $\sigma_{\log{M}}$. 
They find a mean halo mass of $M_{180b} = 13.45 \pm 0.03$ (equivalently 13.43 in our units, $M_{200b}$) and satellite fraction
of $10 \pm 2$\%. Similarly, fitting
to the DR10
projected correlation function and number
density, \cite{Alam17} finds best-fit
values of $M_{\textrm{min}} = 13.25$, $M_1 = 14.18$, $\sigma_{\log{M}} = 0.54$, $M_0 = 12.39$, $\alpha = 1.151$ (friends-of-friends halo masses). This yields
a satellite fraction of 10\% and a mean
halo mass of 13.55 ($M_{200b}$).
These parameters
are in good agreement with our results
except for the slightly lower $\sigma_{\log{M}}$.
\cite{Reid14} performs a similar
fit, also including the monopole and quadrupole
and varying the cosmological parameter
$f\sigma_8$. They find $M_{\textrm{min}} = 13.03 \pm 0.03$, $\sigma_{\log M} = 0.38 \pm 0.06$,
$M_0 = 13.27 \pm 0.13$, $M_1 = 14.08 \pm 0.06$,  and $\alpha = 0.76 \pm 0.18$, for a satellite fraction of $10.2 \pm 0.7$\%,
and a mean halo mass of 13.52.

\begin{figure*}
\includegraphics[width=\textwidth]{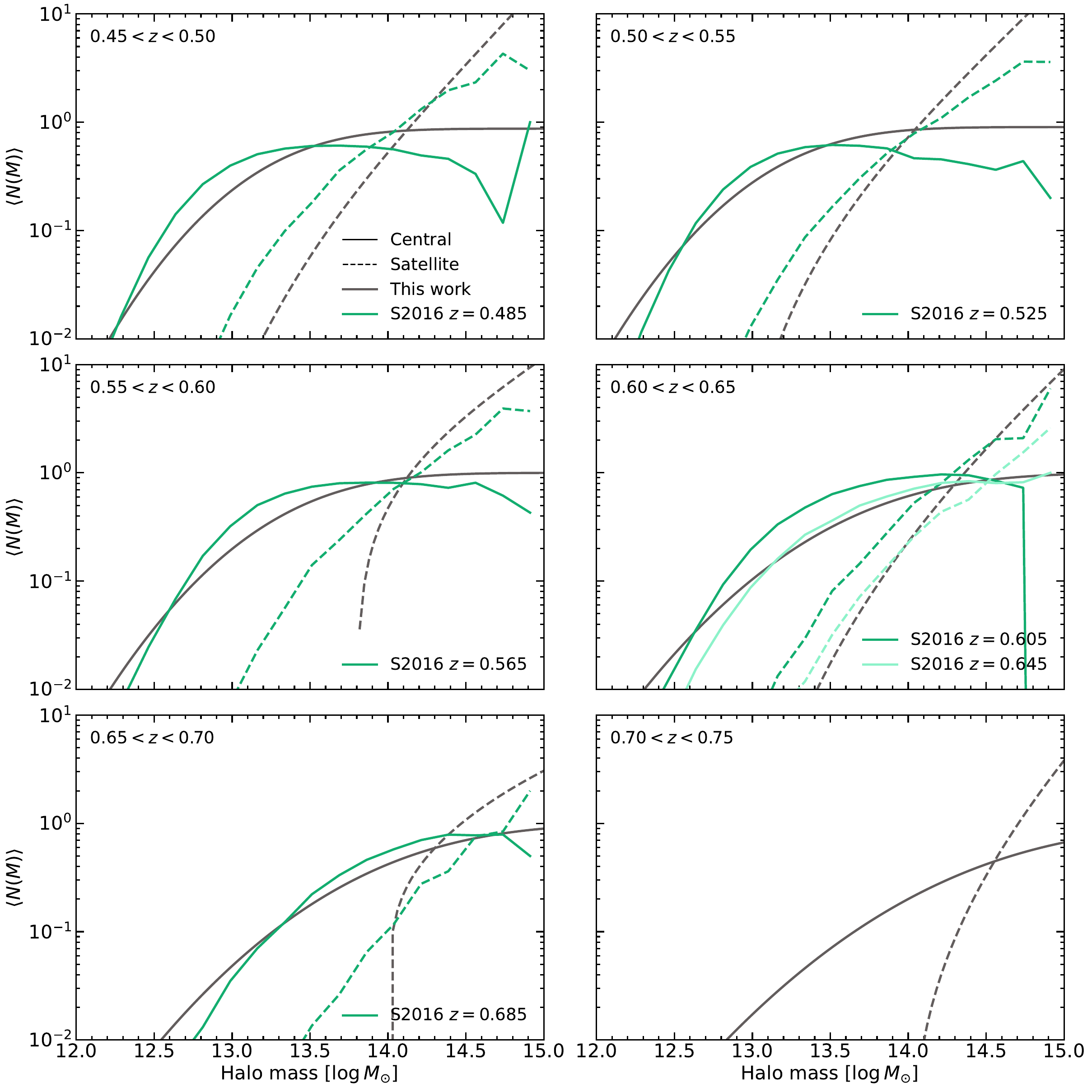}
    \caption{Comparison between our best-fit $\langle N(m) \rangle$ to the CMASS galaxies (gray lines) and those of \citet{Saito16} (S2016; green lines). The solid lines depict $\langle N_{\mathrm{c}}(m) \rangle$ while the dashed lines depict $\langle N_{\mathrm{s}}(m) \rangle$. The S2016 measurements are reported in different redshift bins from our measurements, so we choose the nearest redshift bin to the center of each, reporting their central redshift value in each panel and comparing two bins from S2016 to our $0.6 < z < 0.65$ measurement. Note the qualitative similarities between the results in this paper and those in S2016 despite quantitative differences in methodology. \vspace{1cm}}
    \label{fig:cmass_hod_vs_saito}
\end{figure*}

Going beyond HOD measurements in a single
redshift bin, \cite{Saito16} fit the projected
correlation function (across the entire redshift range)
and the stellar mass function in narrow redshift
bins to create mock catalogs
that match the CMASS $n(z)$ and contain
a redshift-evolving HOD. We compare
our results to their publicly-available
subhalo abundance-matched (SHAM) mock catalogs in Fig.~\ref{fig:cmass_hod_vs_saito}. While
both approaches find a redshift-evolving
HOD, they are quite different
in that \cite{Saito16} only fit to the clustering
in a single bin and use the stellar
mass function to determine the redshift
evolution, whereas we fit to the clustering
in narrow redshift bins. Despite these
differences, our HODs are qualitatively
consistent in shape -- i.e. $N_{\mathrm{c}}(M)$ and $N_{\mathrm{s}}(M)$ match well. However, \cite{Saito16} finds a strong
evolution in mean halo mass, from $\log_{10}(M_{\mathrm{h}}/M_{\odot})$ = 13.12 at $z = 0.445$ to 13.66 at $z = 0.685$, driven by the strong evolution in mean
stellar mass. In contrast, we find a nearly constant mean halo mass.
\cite{Saito16} notes that their SHAM is inconsistent
with the redshift-space monopole
and quadrupole measured across the entire CMASS redshift range.
Their model predicts a strong redshift evolution in the clustering amplitude of the multipoles (due to the strong increase in mean halo mass), which is not observed: the data is nearly constant in redshift (i.e.\ $b(z) D(z)$ is nearly constant).
In contrast, our model is fit to the redshift-dependent
clustering, and hence it shows much less evolution
in mean halo mass and $b(z) D(z)$.
Our results agree better with \cite{Saito16} in the satellite fraction:
they find a drop in the satellite
fraction from 12\% to 7\% between $z = 0.5$
and 0.7, matching our drop from 10\% to 6\%.

\begin{table*}
\begin{tabular}{c|c|c|c|c|c}
    CMASS bin & $M_{\textrm{min}}$ & $M_1$ & $\alpha$ & $M_0$ & $\sigma_{\log{M}}$ \\[5pt]
        \hline
        $0.45 < z < 0.50$ & $13.28^{+0.040}_{-0.041}$ (13.29) &$14.10^{+0.024}_{-0.034}$ (14.11) &$1.50^{+0.077}_{-0.091}$ (1.54) &$12.5^{+0.46}_{-0.36}$ (12.1) &$0.67^{+0.047}_{-0.054}$ (0.67) \\[5pt]
        $0.50 < z < 0.55$ & $13.26^{+0.039}_{-0.037}$ (13.25) &$14.03^{+0.045}_{-0.093}$ (14.01) &$1.39^{+0.115}_{-0.152}$ (1.37) &$12.9^{+0.60}_{-0.59}$ (13.2) &$0.70^{+0.041}_{-0.040}$ (0.69) \\[5pt]
        $0.55 < z < 0.60$ & $13.48^{+0.045}_{-0.038}$ (13.45) &$14.10^{+0.067}_{-0.181}$ (13.84) &$1.28^{+0.186}_{-0.274}$ (0.94) &$13.2^{+0.54}_{-0.80}$ (13.8) &$0.78^{+0.042}_{-0.036}$ (0.75) \\[5pt]
        $0.60 < z < 0.65$ & $13.84^{+0.059}_{-0.047}$ (13.82) &$14.22^{+0.068}_{-0.129}$ (14.27) &$1.11^{+0.285}_{-0.287}$ (1.34) &$13.0^{+0.69}_{-0.66}$ (13.0) &$0.93^{+0.042}_{-0.036}$ (0.92) \\[5pt]
        $0.65 < z < 0.70$ & $14.17^{+0.069}_{-0.062}$ (14.14) &$14.27^{+0.124}_{-0.299}$ (14.00) &$0.77^{+0.467}_{-0.457}$ (0.56) &$13.7^{+0.35}_{-1.05}$ (14.0) &$0.99^{+0.046}_{-0.045}$ (0.97) \\[5pt]
        $0.70 < z < 0.75$ & $14.69^{+0.114}_{-0.106}$ (14.65) &$14.54^{+0.159}_{-0.317}$ (14.41) &$0.98^{+0.853}_{-0.634}$ (1.39) &$13.5^{+0.73}_{-1.01}$ (14.0) &$1.13^{+0.063}_{-0.068}$ (1.11) \\[5pt]
        \hline
Derived parameters & Bias & $\bar{M_{\mathrm{h}}}$ ($\log{M_{\odot}}$) & $f_{\textrm{sat}}$ & $\chi^2$ \\[5pt]
        \cline{1-5}
        $0.45 < z < 0.50$ & $1.88^{+0.025}_{-0.022}$ (1.88) &$13.611^{+0.008}_{-0.008}$ (13.611) &$0.107^{+0.009}_{-0.009}$ (0.110) & 16.155 \\[5pt]
        $0.50 < z < 0.55$ & $1.87^{+0.017}_{-0.018}$ (1.87) &$13.555^{+0.006}_{-0.007}$ (13.555) &$0.099^{+0.018}_{-0.020}$ (0.091)& 14.498  \\[5pt]
        $0.55 < z < 0.60$ & $1.96^{+0.019}_{-0.021}$ (1.97) &$13.551^{+0.008}_{-0.008}$ (13.556) &$0.078^{+0.022}_{-0.017}$ (0.059) & 6.675 \\[5pt]
        $0.60 < z < 0.65$ & $2.02^{+0.022}_{-0.022}$ (2.02) &$13.543^{+0.009}_{-0.010}$ (13.547) &$0.073^{+0.034}_{-0.020}$ (0.069) & 9.033 \\[5pt]
        $0.65 < z < 0.70$ & $2.21^{+0.036}_{-0.037}$ (2.22) &$13.597^{+0.014}_{-0.015}$ (13.602) &$0.059^{+0.033}_{-0.017}$ (0.046) & 9.012 \\[5pt]
$0.70 < z < 0.75$ & $2.30^{+0.069}_{-0.054}$ (2.31) &$13.602^{+0.025}_{-0.025}$ (13.610) &$0.034^{+0.050}_{-0.018}$ (0.021) & 5.772 \\[5pt]
\end{tabular}
    \caption{Marginalized and best-fit parameters for CMASS galaxies, requiring unWISE-Blue HOD to match observed number density. $\chi^2$ is the combined $\chi^2$ for the entire data vector (9 degrees of freedom).}
    \label{tab:results_cmass}
\end{table*}

\clearpage

\begin{table*}
\begin{tabular}{c|c|c|c|c|c}
    unWISE-Blue bin & $M_{\textrm{min}}$ & $M_1$ & $\alpha$ & $M_0$ & $\sigma_{\log{M}}$ \\[5pt]
        \hline
    $0.45 < z < 0.50$ & $12.20^{+0.083}_{-0.051}$ (12.16) &$13.16^{+0.137}_{-0.208}$ (13.09) &$1.11^{+0.132}_{-0.153}$ (1.06) &$12.9^{+0.27}_{-0.51}$ (13.0) &$0.23^{+0.213}_{-0.141}$ (0.07)\\[5pt]
    $0.50 < z < 0.55$ & $12.28^{+0.062}_{-0.048}$ (12.24) &$13.33^{+0.049}_{-0.097}$ (13.30) &$1.30^{+0.046}_{-0.104}$ (1.28) &$12.2^{+0.50}_{-0.48}$ (12.4) &$0.22^{+0.161}_{-0.107}$ (0.15) \\[5pt]
 $0.55 < z < 0.60$ & $12.66^{+0.244}_{-0.206}$ (12.73) &$13.35^{+0.069}_{-0.165}$ (13.27) &$1.19^{+0.073}_{-0.166}$ (1.10) &$12.4^{+0.52}_{-0.52}$ (12.8) &$0.87^{+0.217}_{-0.227}$ (0.96) \\[5pt]
$0.60 < z < 0.65$ & $12.99^{+0.238}_{-0.213}$ (13.00) &$13.23^{+0.181}_{-0.260}$ (13.11) &$0.89^{+0.177}_{-0.177}$ (0.82) &$13.0^{+0.23}_{-0.41}$ (13.1) &$1.15^{+0.162}_{-0.173}$ (1.18) \\[5pt]
$0.65 < z < 0.70$ & $12.90^{+0.260}_{-0.238}$ (12.95) &$13.50^{+0.064}_{-0.087}$ (13.51) &$1.15^{+0.089}_{-0.119}$ (1.16) &$12.2^{+0.55}_{-0.47}$ (11.7) &$1.04^{+0.195}_{-0.219}$ (1.08) \\[5pt]
$0.70 < z < 0.75$ & $12.53^{+0.281}_{-0.175}$ (12.39) &$13.38^{+0.152}_{-0.486}$ (12.47) &$1.14^{+0.219}_{-0.366}$ (0.62) &$13.0^{+0.43}_{-0.87}$ (13.6) &$0.61^{+0.309}_{-0.362}$ (0.38) \\[5pt]
    \hline
Derived parameters & Bias & $\bar{M_{\mathrm{h}}}$ ($\log{M_{\odot}}$) & $f_{\textrm{sat}}$\\[5pt]
        \cline{1-4}
$0.45 < z < 0.50$ & $1.53^{+0.020}_{-0.025}$ (1.54) &$13.404^{+0.012}_{-0.013}$ (13.407) &$0.185^{+0.023}_{-0.013}$ (0.180) \\[5pt]
 $0.50 < z < 0.55$ & $1.63^{+0.019}_{-0.020}$ (1.63) &$13.430^{+0.011}_{-0.012}$ (13.430) &$0.220^{+0.021}_{-0.022}$ (0.216) \\[5pt]
 $0.55 < z < 0.60$ & $1.48^{+0.054}_{-0.059}$ (1.45) &$13.286^{+0.022}_{-0.023}$ (13.273) &$0.179^{+0.026}_{-0.023}$ (0.164) \\[5pt]
 $0.60 < z < 0.65$ & $1.37^{+0.051}_{-0.048}$ (1.35) &$13.140^{+0.024}_{-0.023}$ (13.131) &$0.132^{+0.020}_{-0.013}$ (0.127) \\[5pt]
$0.65 < z < 0.70$ & $1.43^{+0.068}_{-0.064}$ (1.41) &$13.114^{+0.031}_{-0.030}$ (13.104) &$0.143^{+0.031}_{-0.028}$ (0.184) \\[5pt]
$0.70 < z < 0.75$ & $1.64^{+0.074}_{-0.095}$ (1.69) &$13.199^{+0.036}_{-0.037}$ (13.223) &$0.110^{+0.038}_{-0.022}$ (0.088) \\[5pt]
\end{tabular}
    \caption{Marginalized and best-fit parameters for unWISE galaxies, requiring unWISE-Blue HOD to match observed number density. $\chi^2$ is not shown, as Table~\ref{tab:results_cmass} shows the combined $\chi^2$.
    \label{tab:results_wise}}
\end{table*}

\subsection{unWISE-Blue}

\subsubsection{Redshift evolution of the unWISE-Blue HOD---baseline results}

\begin{figure*}
    \centering
    \includegraphics[width=1.0\textwidth]{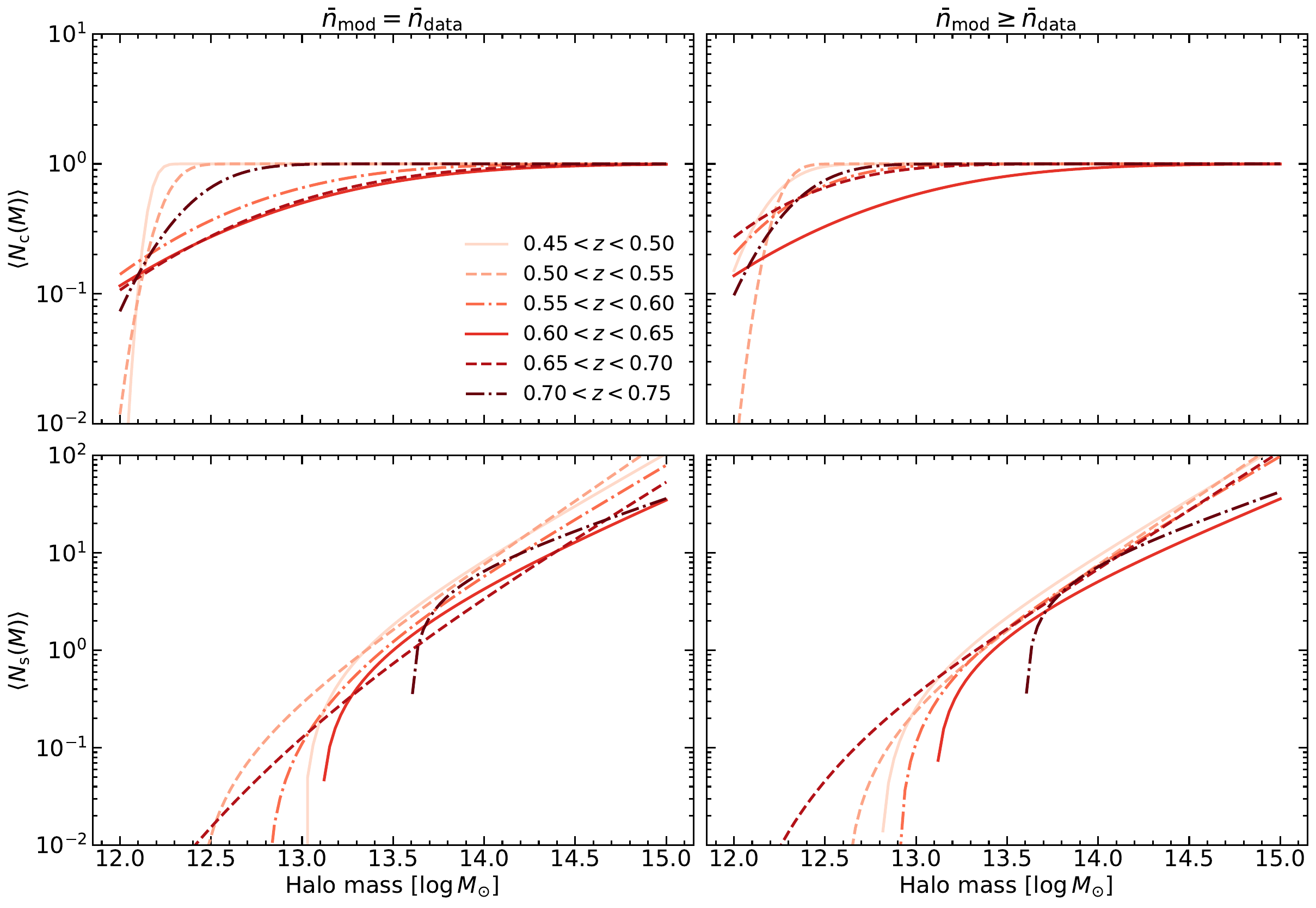}
    \caption{The mean number of central galaxies (top row) and the mean number of satellite galaxies (bottom row) for the unWISE-Blue HOD as functions of halo mass. The left column corresponds to the default case where we use the observed value of $\bar{n}$, while the right column corresponds to the ``incompleteness" case where the model is allowed to have a higher $\bar{n}$ than the data. Plotting all redshift bins demonstrates the evolution of the unWISE-Blue HOD as a function of redshift. 
    As in Fig.~\ref{fig:cmass_hod_evolution}, the sharp drops in $\langle N_{\mathrm{s}}(M) \rangle$ occurr when the halo mass drops below $M_0$.}
    \label{fig:unwise_hod_evolution}    
\end{figure*}

Like CMASS, the unWISE-Blue sample HOD also evolves with redshift.
If we require the model to match the unWISE number density (i.e.\ not allowing for incompleteness in $\langle N_{\mathrm{c}}(M) \rangle$), the largest
difference is the dramatic increase of $M_{\textrm{min}}$
and $\sigma_{\log{M}}$ from the $0.5 < z < 0.55$ bin to $z > 0.55$. 
This leads to a considerably different $\langle N_{\textrm{c}}(M) \rangle$ at $0.5 < z < 0.55$ than
in the higher redshift bins (Fig.~\ref{fig:unwise_hod_evolution}).
We note that $M_{\textrm{min}}$
and $\sigma_{\log{M}}$ are strongly degenerate; while
the 1D posteriors are very far apart on both parameters
comparing $0.50 < z < 0.55$ and $0.55 < z < 0.60$, the 2D posteriors in the $M_{\textrm{min}}$--$\sigma_{\log{M}}$ plane just barely touch.

As with CMASS, we find that the best-fit HOD
from the other redshift bins is strongly ruled out by the data
in any redshift bin. We find that the average halo mass
and bias generally decreases with redshift for the unWISE samples, with $\log_{10}(M_{\mathrm{h}}/M_{\odot})$ dropping
from
13.4 at $z \sim 0.5$ to 13.1 at $z \sim 0.65$, and the bias dropping from 1.6 to 1.4 over the same redshift range.
This is in contrast to the rough fit of \cite{Krolewski2019}, who
find $b(z) = 0.8 + 1.2z$, implying an increase from $b = 1.43$ to $b = 1.61$ from $z = 0.5$ to 0.7.
However, that fitting function was a very approximate
fit to underlying bias measurements
that also decreased across the redshift range considered here of $0.45 < z < 0.75$,
from 1.75 to 1.68 to 1.59 to 1.58 (dip in left panel of Fig.\ 19 in \cite{Krolewski2019}). The overall increasing $b(z)$ was intended to match the behavior over a much wider redshift range, $0 < z < 1$.
Compared to CMASS, the unWISE galaxies have lower halo mass ($\log_{10}(M_{\mathrm{h}}/M_{\odot})$ $\sim 13.25$ vs. 13.55) and higher satellite fraction (15--20\% vs.\ 6--10\%).

\subsubsection{Allowing for incompleteness in unWISE-Blue}

Tables~\ref{tab:results_cmass_wise_incompleteness} and~\ref{tab:results_wise_incompleteness} show our results if we allow for arbitrary
incompleteness in the unWISE-Blue HOD. The unWISE-Blue HOD parameters
are affected most strongly at $z > 0.55$. This is because the clustering information alone favors an HOD matching the observed
$\bar{n}(z)$ in the first two bins, whereas it prefers a larger $\bar{n}(z)$ at higher redshift.
That is, the best-fit $f_{\textrm{inc}}$ is around 1 in the first two bins, but drops to around 0.5 at higher redshift. The most notable change at higher-$z$ is therefore a shift along the $M_{\textrm{min}}$--$\sigma_{\log{M}}$
degeneracy towards lower values of $M_{\textrm{min}}$ and higher number density. This shift is apparent in Fig.~\ref{fig:unwise_hod_evolution}: $N_{\textrm{c}}$ with arbitrary incompleteness extends to lower halo mass, with a very slightly steeper dropoff, though not nearly as sharp as $N_{\textrm{c}}$ in the first two redshift bins.

When allowing for unWISE incompleteness, the CMASS parameters do not shift much, though the uncertainties are somewhat larger. The derived parameters are more stable to this change: the CMASS bias and mean halo mass (and their uncertainties) are nearly unchanged, and the median satellite fraction is similar but with somewhat larger errorbars. Likewise, the unWISE-Blue HODs have small changes in bias, mean halo mass, and satellite fraction,
but these changes are all within the errorbar.

\subsubsection{Comparison to past work}

\cite{Kusiak22} performs a similar analysis to us, fitting an HOD to the small-scale unWISE angular auto-correlation ($C_{\ell}^{gg}$) and cross-correlation with CMB lensing ($C_{\ell}^{\kappa g}$). Unlike this work,
their HOD is redshift-independent, as they do not perform a tomographic measurement but rather consider only the galaxy auto-correlation and the cross-correlation with the broad CMB lensing kernel, neither of which allow them to extract redshift information. They also use a slightly different HOD model, fixing $M_0$ to zero and allowing the NFW truncation radius to vary as a free parameter.

For the unWISE-Blue sample, they find $\log_{10}(M_{\textrm{min}}/M_{\odot}) = 12.11\pm0.37$, somewhat lower than our values. They also find a lower value for $M_1$, a slightly higher value for $\alpha$, and $\sigma_{\log{M}} = 0.73^{+0.33}_{-0.22}$. Their derived parameters are quite consistent with ours,
with a mean halo mass of $\log_{10}(M_{\mathrm{h}}/M_{\odot}) = 13.26 \pm 0.02$, $b = 1.50^{+0.01}_{-0.02}$, and satellite fraction of $0.23^{+0.06}_{-0.07}$.
Our results suggest that while this redshift-independent HOD can fit the clustering of the entire unWISE-Blue sample well, a redshift-evolving HOD (with similar parameters on average) better fits the tomographic cross-correlation with CMASS spectroscopic galaxies. Our approach allows for dramatically improved resolution on the redshift evolution of the unWISE-Blue HOD, with better redshift resolution than previous work on different samples (i.e.\ the $\Delta z \sim 0.15$ photometric redshift bins in the DES redMaGiC and MagLim HOD analysis of \cite{Zacharegkas22}).

\section{Conclusions}  \label{sec:conclusions}

We have measured the small-scale cross-correlation
between the unWISE-Blue galaxies, selected via an infrared color cut from WISE imaging and possessing a broad redshift distribution at $0 < z < 1$, with the BOSS CMASS spectroscopic sample in narrow redshift bins of width $\Delta z =0.05$ between $z = 0.45$ and 0.75. This allowed us to tomographically probe the unWISE and CMASS halo occupation distribution. We augmented the halo model, modifying the \texttt{HALOMOD} code \citep{Halomod20}, with the tools necessary to model this cross-correlation,
including redshift-space distortions,
beyond-Limber corrections, and halo exclusion. Our model fits the CMASS-unWISE cross-correlation and CMASS autocorrelation well at $0.1 < r_p < 10$ $h^{-1}$ Mpc.

We find that the CMASS HOD evolves
strongly at $0.45 < z < 0.75$, with $\log_{10}(M_{\textrm{min}}/M_{\odot})$ increasing from 13.28 to 14.67 and the scatter $\sigma_{\log{M}}$ increasing from 0.67 to 1.11. The strong evolution in the halo mass function opposes the change in the HOD to yield a mean halo mass that is nearly constant, around $\log_{10}(M_{\mathrm{h}}/M_{\odot}) = 13.55$.
The mean bias increases significantly, from 1.88 at $z = 0.475$ to 2.32 at $z = 0.725$. These results are largely consistent with past results, which generally fit the clustering of the sample across the entire redshift range \citep{White2011,Reid14,Saito16,Alam17}.\footnote{There have been a large number of works studying the CMASS HOD \citep{Guo13,Guo14,Favole16,Zhai17,Yuan20,Yuan21,Yuan22,Zhai23}, but we focus on comparing to \citet{White2011}, \citet{Reid14}, \citet{Saito16}, and \citet{Alam17} as the others apply additional cuts to the sample, e.g. in redshift and luminosity.}

The unWISE-Blue HOD is more constant than the CMASS HOD, and the evolution of the HOD parameters depends on whether we assume the HOD is complete at high masses, or allow for $\langle N_c(M) \rangle$ to asymptote to a value less than one at high halo masses. The evolution of derived parameters (mean bias, mean halo mass, and satellite fraction) is similar regardless of whether we allow for high-mass incompleteness. The bias is much more constant for unWISE-Blue than for CMASS; indeed, it declines only from $\sim$1.6 to $\sim$1.4 between $z\sim0.5$ and 0.7.
Likewise, the mean halo mass declines slightly from $\log_{10}(M_{\mathrm{h}}/M_{\odot} \sim 13.4$ to $\sim$13.1 from $z = 0.5$ to 0.7. The satellite fraction of $\sim$20\% is higher than the satellite fraction in BOSS, and declines slightly towards higher redshift.

The tomographic measurement of the unWISE-Blue HOD complements the previous unWISE-Blue HOD measurement of \cite{Kusiak22} from the angular auto-correlation and the CMB lensing cross-correlation. Our HODs broadly agree
with those of \cite{Kusiak22}, but our approach
has the ability to test whether the HOD is redshift-independent, and indeed
favors a modest evolution at $0.45 < z < 0.75$, spanning the peak of the unWISE-Blue redshift distribution. This distinction is potentially significant given the broad redshift distribution,
and its implications for the unWISE auto-correlation and CMB lensing cross-correlation
will be explored in future work.\footnote{Due to the broad redshift distribution, predicting these statistics with our tomographic HOD model requires constraints on the HOD at $z < 0.45$ and $z > 0.75$, i.e.\ from cross-correlations with other spectroscopic samples such as BOSS-LOWZ and eBOSS LRGs.}
Our tomographic HOD model will also be useful for constructing mock catalogs for the unWISE-Blue sample (which are essential for cosmological modelling \citep{Krolewski2021,Farren23})
and for further enabling the use of this sample in cross-correlation measurements.

\section*{Acknowledgments}

AK was supported as a CITA National Fellow by the Natural Sciences and Engineering Research Council of Canada (NSERC), funding reference \#DIS-2022-568580.
JL acknowledges receipt of a Natural Sciences and Engineering Research Council of Canada (NSERC) Undergraduate Student Research Award. WP acknowledges support from NSERC [funding reference number RGPIN-2019-03908] and from the Canadian Space Agency. Research at Perimeter Institute is supported in part by the Government of Canada through the Department of Innovation, Science and Economic Development Canada and by the Province of Ontario through the Ministry of Colleges and Universities. This research was enabled in part by support provided by Compute Ontario (\hyperlink{computeontario.ca}{computeontario}) and the Digital Research Alliance of Canada (\hyperlink{alliancecan.ca}{alliancecan}).


\section*{Code Availability}
The code used to fit the data is available at \url{https://github.com/jensenlawrence/CMASS-WISE-HOD}.

\begin{acknowledgements}
      
\end{acknowledgements}

%
%

\bibliographystyle{aa.bst}
\bibliography{refs.bib}

\appendix

\begin{figure*}
\includegraphics[width=\textwidth]{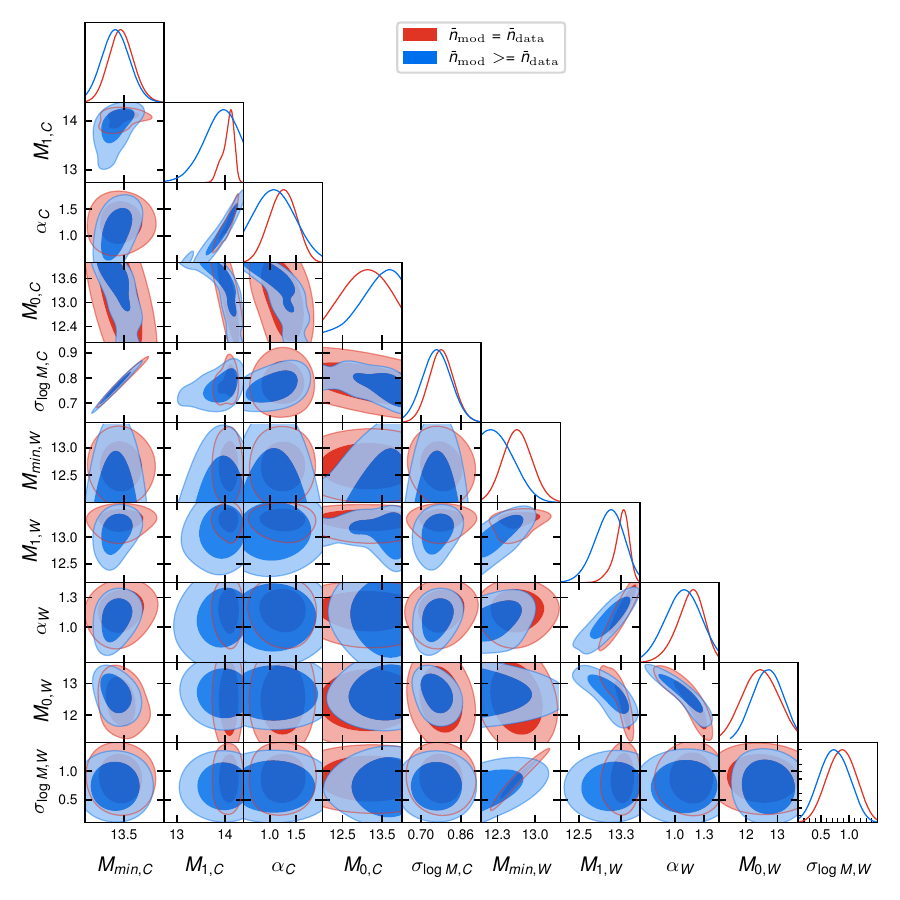}
    \caption{Constraints and degeneracies between the 10 HOD parameters at $0.55 < z < 0.6$. A subscript $C$ denotes the CMASS HOD parameters while a subscript $W$ denotes the unWISE-Blue HOD parameters. The default model is in red, and the model allowing for unWISE incompleteness is in blue. Parameter degeneracies are similar in other redshift bins.}
    \label{fig:contours}
\end{figure*}

\begin{figure*}
\includegraphics[width=\textwidth]{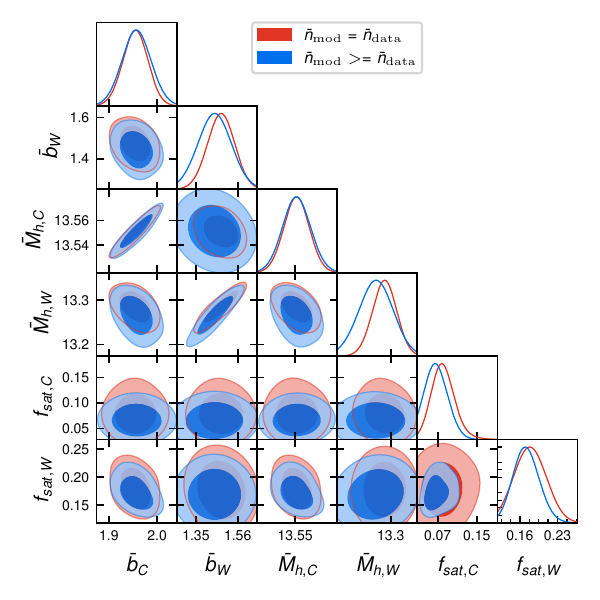}
    \caption{Constraints and degeneracies between the derived parameters at $0.55 < z < 0.6$. A subscript $C$ denotes the CMASS HOD parameters while a subscript $W$ denotes the unWISE-Blue HOD parameters. The default model is in red, and the model allowing for unWISE incompleteness is in blue.}
    \label{fig:contours_derived}
\end{figure*}

Here we present additional tables and plots. In Tables~\ref{tab:results_cmass_wise_incompleteness} and~\ref{tab:results_wise_incompleteness}, we give the marginalized constraints on the HOD and derived parameters for both CMASS and unWISE-Blue, allowing the unWISE-Blue model number density to exceed the data number density (i.e.\ allowing for an arbitrary $f_{\textrm{inc}}$ by removing the unWISE-Blue number density from the likelihood). In Figs.~\ref{fig:contours} and~\ref{fig:contours_derived}, we show
contour plots for both the HOD and derived parameters for a representative redshift bin, $0.55 < z < 0.6$.

\FloatBarrier



\begin{table*}
\begin{tabular}{c|c|c|c|c|c}
    CMASS bin & $M_{\textrm{min}}$ & $M_1$ & $\alpha$ & $M_0$ & $\sigma_{\log{M}}$ \\[5pt]
        \hline
        $0.45 < z < 0.50$ & $13.28^{+0.038}_{-0.039}$ (13.29) &$14.10^{+0.023}_{-0.030}$ (14.11) &$1.49^{+0.072}_{-0.081}$ (1.54) &$12.5^{+0.42}_{-0.36}$ (12.1) &$0.67^{+0.045}_{-0.050}$ (0.68) \\[5pt]
$0.50 < z < 0.55$ & $13.26^{+0.042}_{-0.038}$ (13.23) &$14.02^{+0.048}_{-0.079}$ (13.99) &$1.37^{+0.119}_{-0.131}$ (1.36) &$12.9^{+0.50}_{-0.60}$ (13.4) &$0.70^{+0.043}_{-0.041}$ (0.66) \\[5pt]
$0.55 < z < 0.60$ & $13.46^{+0.046}_{-0.037}$ (13.43) &$13.96^{+0.181}_{-0.310}$ (13.23) &$1.07^{+0.303}_{-0.356}$ (0.49) &$13.7^{+0.24}_{-0.77}$ (14.0) &$0.76^{+0.042}_{-0.038}$ (0.73) \\[5pt]
$0.60 < z < 0.65$ & $13.86^{+0.064}_{-0.051}$ (13.83) &$14.23^{+0.066}_{-0.084}$ (14.25) &$1.13^{+0.278}_{-0.274}$ (1.23) &$12.8^{+0.67}_{-0.52}$ (13.0) &$0.94^{+0.043}_{-0.038}$ (0.92)\\[5pt]
$0.65 < z < 0.70$ & $14.16^{+0.066}_{-0.058}$ (14.14) &$14.19^{+0.193}_{-0.828}$ (13.77) &$0.66^{+0.567}_{-0.463}$ (0.33) &$13.8^{+0.30}_{-1.09}$ (14.1) &$0.98^{+0.043}_{-0.042}$ (0.97) \\[5pt]
$0.70 < z < 0.75$ & $14.63^{+0.091}_{-0.087}$ (14.63) &$12.78^{+0.503}_{-0.372}$ (12.61) &$0.24^{+0.295}_{-0.177}$ (0.32) &$14.4^{+0.22}_{-0.19}$ (14.6) &$1.10^{+0.057}_{-0.058}$ (1.10) \\[5pt]
        \hline
Derived parameters & Bias & $\bar{M_{\mathrm{h}}}$ ($\log{M_{\odot}}$) & $f_{\textrm{sat}}$ & $\chi^2$\\[5pt]
\cline{1-5}
$0.45 < z < 0.50$ & $1.88^{+0.021}_{-0.021}$ (1.88) &$13.610^{+0.008}_{-0.007}$ (13.610) &$0.108^{+0.009}_{-0.009}$ (0.110) & 16.403  \\[5pt]
$0.50 < z < 0.55$ & $1.87^{+0.016}_{-0.017}$ (1.88) &$13.555^{+0.006}_{-0.006}$ (13.558) &$0.098^{+0.020}_{-0.018}$ (0.083) & 14.372 \\[5pt]
$0.55 < z < 0.60$ & $1.96^{+0.021}_{-0.021}$ (1.96) &$13.551^{+0.009}_{-0.008}$ (13.549) &$0.064^{+0.022}_{-0.011}$ (0.052) & 6.651 \\[5pt]
$0.60 < z < 0.65$ & $2.02^{+0.020}_{-0.023}$ (2.03) &$13.542^{+0.009}_{-0.009}$ (13.547) &$0.081^{+0.037}_{-0.022}$ (0.074) & 9.089 \\[5pt]
$0.65 < z < 0.70$ & $2.21^{+0.035}_{-0.034}$ (2.22) &$13.597^{+0.014}_{-0.014}$ (13.599) &$0.053^{+0.029}_{-0.017}$ (0.044) & 9.042 \\[5pt]
$0.70 < z < 0.75$ & $2.31^{+0.066}_{-0.055}$ (2.30) &$13.604^{+0.024}_{-0.023}$ (13.604) &$0.015^{+0.012}_{-0.008}$ (0.012) & 5.730 \\[5pt]
 
\end{tabular}
    \caption{As in Table~\ref{tab:results_cmass}, but for the case where the model number density is allowed to exceed the observed number density.
    \label{tab:results_cmass_wise_incompleteness}}
\end{table*}

\begin{table*}
\begin{tabular}{c|c|c|c|c|c}
    unWISE-Blue bin & $M_{\textrm{min}}$ & $M_1$ & $\alpha$ & $M_0$ & $\sigma_{\log{M}}$ \\[5pt]
        \hline
$0.45 < z < 0.50$ & $12.19^{+0.093}_{-0.082}$ (12.19) &$13.14^{+0.147}_{-0.213}$ (13.10) &$1.11^{+0.138}_{-0.153}$ (1.06) &$12.8^{+0.28}_{-0.64}$ (13.0) &$0.26^{+0.210}_{-0.168}$ (0.22) \\[5pt]
$0.50 < z < 0.55$ & $12.26^{+0.066}_{-0.064}$ (12.24) &$13.31^{+0.059}_{-0.092}$ (13.28) &$1.30^{+0.049}_{-0.093}$ (1.25) &$12.3^{+0.44}_{-0.49}$ (12.6) &$0.21^{+0.160}_{-0.110}$ (0.13) \\[5pt]
$0.55 < z < 0.60$ & $12.38^{+0.287}_{-0.243}$ (12.32) &$13.11^{+0.200}_{-0.312}$ (13.17) &$1.10^{+0.115}_{-0.177}$ (1.09) &$12.7^{+0.34}_{-0.50}$ (12.9) &$0.75^{+0.195}_{-0.187}$ (0.54) \\[5pt]
$0.60 < z < 0.65$ & $12.40^{+0.481}_{-0.510}$ (12.84) &$12.98^{+0.268}_{-0.298}$ (13.00) &$0.98^{+0.144}_{-0.190}$ (0.78) &$12.8^{+0.34}_{-0.71}$ (13.1) &$0.87^{+0.283}_{-0.396}$ (1.09) \\[5pt]
$0.65 < z < 0.70$ & $12.29^{+0.529}_{-0.382}$ (12.30) &$13.20^{+0.235}_{-0.238}$ (13.30) &$1.15^{+0.088}_{-0.117}$ (1.20) &$12.3^{+0.53}_{-0.51}$ (12.0) &$0.73^{+0.352}_{-0.334}$ (0.70) \\[5pt]
$0.70 < z < 0.75$ & $12.34^{+0.265}_{-0.256}$ (12.33) &$13.10^{+0.280}_{-0.431}$ (12.44) &$1.05^{+0.227}_{-0.273}$ (0.64) &$13.1^{+0.34}_{-0.62}$ (13.6) &$0.49^{+0.342}_{-0.309}$ (0.36) \\[5pt]
    \hline
Derived parameters & Bias & $\bar{M_{\mathrm{h}}}$ ($\log{M_{\odot}}$) & $f_{\textrm{sat}}$ & $f_{\textrm{inc}}$\\[5pt]
        \cline{1-5}
$0.45 < z < 0.50$ & $1.52^{+0.026}_{-0.034}$ (1.54) &$13.400^{+0.013}_{-0.016}$ (13.408) &$0.187^{+0.030}_{-0.014}$ (0.180) &$0.830^{+0.154}_{-0.083}$ (0.874)\\[5pt]
$0.50 < z < 0.55$ & $1.62^{+0.022}_{-0.026}$ (1.63) &$13.427^{+0.012}_{-0.013}$ (13.430) &$0.220^{+0.021}_{-0.020}$ (0.208) &$0.949^{+0.132}_{-0.091}$ (0.960)   \\[5pt]
$0.55 < z < 0.60$ & $1.44^{+0.059}_{-0.053}$ (1.52) &$13.266^{+0.028}_{-0.028}$ (13.299) &$0.169^{+0.024}_{-0.016}$ (0.157) &$0.480^{+0.241}_{-0.127}$ (0.563)  \\[5pt]
$0.60 < z < 0.65$ & $1.35^{+0.053}_{-0.044}$ (1.36) &$13.118^{+0.026}_{-0.026}$ (13.126) &$0.145^{+0.045}_{-0.019}$ (0.126) &$0.478^{+0.394}_{-0.177}$ (0.780)  \\[5pt]
$0.65 < z < 0.70$ & $1.39^{+0.066}_{-0.066}$ (1.42) &$13.082^{+0.033}_{-0.037}$ (13.094) &$0.144^{+0.035}_{-0.028}$ (0.155) &$0.619^{+0.445}_{-0.252}$ (0.648) \\[5pt]
$0.70 < z < 0.75$ & $1.57^{+0.093}_{-0.105}$ (1.67) &$13.170^{+0.045}_{-0.050}$ (13.212) &$0.104^{+0.034}_{-0.019}$ (0.091) &$0.738^{+0.423}_{-0.195}$ (0.895)  \\[5pt]
\end{tabular}
    \caption{As in Table~\ref{tab:results_wise}, but for the case where the model number density is allowed to exceed the observed number density.}
    \label{tab:results_wise_incompleteness}
\end{table*}

\FloatBarrier

\FloatBarrier

\end{document}